\documentstyle[11pt,epsf]{article}

\input psfig
\def\beq{\begin{eqnarray}}
\def\eeq{\end{eqnarray}}
\def\bea{\begin{eqnarray*}}
\def\eea{\end{eqnarray*}}

\def\centeron#1#2{{\setbox0=\hbox{#1}\setbox1=\hbox{#2}\ifdim
\wd1>\wd0\kern.5\wd1\kern-.5\wd0\fi
\copy0\kern-.5\wd0\kern-.5\wd1\copy1\ifdim\wd0>\wd1
\kern.5\wd0\kern-.5\wd1\fi}}
\def\ltap{\;\centeron{\raise.35ex\hbox{$<$}}{\lower.65ex\hbox{$\sim$}}\;}
\def\gtap{\;\centeron{\raise.35ex\hbox{$>$}}{\lower.65ex\hbox{$\sim$}}\;}

\def\singleandthirdspaced{\baselineskip=\normalbaselineskip\multiply
    \baselineskip by 130\divide\baselineskip by 100}
\def\singlespaced{\baselineskip=\normalbaselineskip}

\newcommand{\newc}{\newcommand}
\newc{\qbar}{{\overline q}}
\newc{\Kahler}{K\"ahler }
\newc{\deltaGS}{\delta_{\rm GS}}
\begin{document}
\begin{titlepage}
\begin{flushright}
{\large hep-th/0003175 \\ SCIPP-99/54\\

PRELIMINARY}
\end{flushright}

\vskip 1.2cm

\begin{center}

{\LARGE\bf
TASI Lectures on M Theory Phenomenology}

\vskip 1.4cm

{\large Michael Dine}
\\
\vskip 0.4cm
%{\it $^a$Stanford Linear Accelerator Center,
%     Stanford CA 94309} \\
{\it Santa Cruz Institute for Particle Physics,
     Santa Cruz CA 95064  } \\
%{\it $^c$Physics Department,
%     University of California,
%     Santa Cruz CA 95064  }

\vskip 4pt

\vskip 1.5cm

\begin{abstract}
These lectures discuss some of the general issues in developing a
phenomenology for Superstring Theory/M Theory.  The focus is on
the question:  how might one obtain robust, generic predictions.
For example, does the theory predict low energy supersymmetry
breaking?  In the course of these explorations, basics of
supersymmetry and supersymmetry breaking, string moduli,
cosmological issues, and other questions are addressed.  The
notion of approximate moduli and their possible role plays
a central role in the discussion.

\end{abstract}

\end{center}

\vskip 1.0 cm

\end{titlepage}
\setcounter{footnote}{0} \setcounter{page}{2}
\setcounter{section}{0} \setcounter{subsection}{0}
\setcounter{subsubsection}{0}

%%%%%%%%%%%%%%%%%%%%%%%%%%%%%%%%%%%%%%%%%%%
%%%%%%%%%%%%%%%%%%%%%%%%%%%%
\singleandthirdspaced

%\begin{document}

\section{Introduction}

How might string/M theory make contact with nature?  In one view,
we might imagine that some day, one will find some ``true ground
state," or set of ground states, and that we will simply calculate
all of the quantities of low energy physics.  Another viewpoint
holds that these problems are impossibly hard, at least with any
theoretical tools we have today or which we can see on the
horizon, and that we should focus exclusively on theoretical
issues: theoretical consistency, problems of black holes and other
quantum gravity questions, and the like.

In these lectures, I advocate a middle ground.  While I don't
believe it is likely that we will succeed in solving the theory
completely any time soon, I believe it might be possible
to make a few robust, qualitative statements, and perhaps a
small number of quantitative ones.  If we could reliably assert,
for example (preferably before its discovery), that low energy
supersymmetry is a prediction of string theory, with some rough
pattern of soft breakings, this would be a triumph.  If we could
predict one or two mass ratios, or the value of the gauge
couplings, this would be spectacular.  If string theory resolved
some of the problems of cosmology, this would be a major
achievement.

In these lectures I will not succeed in accomplishing any of these
goals, but I do hope to outline the major issues in bringing
string theory into contact with nature.  Our strategy will be to
focus on the major issues in developing a string phenomenology:
\begin{itemize}
\item
The cosmological constant
\item
The problem of vacuum degeneracy
\item
The Hierarchy problem
\item
The role of supersymmetry
\item
The smallness of the gauge couplings and gauge coupling
unification
\item
The size (and shape) of extra dimensions
\item
The question of CP violation and the strong CP problem
\item
Issues of flavor
\item
Questions in Cosmology
\end{itemize}

In line with our remarks above, the goals of a superstring
phenomenology should be to obtain qualitative, generic
predictions, such as
\begin{itemize}
\item
Low energy supersymmetry
\item
Extra light particles
\item
The pattern of supersymmetry breaking
\item
Statements about early universe cosmology, including dark matter,
the evolution of moduli, possible inflaton candidates, and perhaps
the value of the cosmological constant.
\item
Axions
\item
Predictions for rare processes
\end{itemize}

To appreciate the difficulties, consider the problem of low energy
supersymmetry.  Low energy supersymmetry has long been
touted as a possible solution of the hierarchy problem.
Supersymmetry seems to play a fundamental role in string theory,
and numerous solutions to the classical string equations with $N=1$ supersymmetry
in four dimensions are known.    Still, we have no
reliable computation of a stable vacuum with broken supersymmetry.
Nor do we know of any principle which suggests that vacua with
approximate supersymmetry are somehow more special than those
without.  These problems are closely tied to our lack of
understanding of the cosmological constant problem.

It is easiest, as we will see,
to study supersymmetric states; the more supersymmetry, the
easier. But this by itself is hardly an argument for
supersymmetry.  We will need some more persuasive
argument if we are to make a statement that string theory does or
does not predict low energy supersymmetry.  An overriding question
should be:  can we make such a statement without
knowledge of the precise ground state?

We might hope to get some insight into these questions by
considering two of the problems posed above:  the cosmological
constant problem and the hierarchy problem.  Both represent
failures of dimensional analysis.  It is possible to add to the
effective action of the standard model the terms:
\beq
\int d^4 x \sqrt{g}( \Lambda + M^2 \vert H \vert^2).
\eeq
Here $\Lambda$ is a quantity with dimensions of $M^4$, and so one
might expect that it is of order some large scale in nature,
$\Lambda \sim M_p^4$.  On the other hand, there are recent
claims\cite{cosmocon} that a cosmological constant has been
observed, with a value of order
\beq
\Lambda \approx 10^{-47}~ {\rm GeV}^4.
\eeq
Even if one is skeptical of this result (and evidence is
steadily mounting that it is correct), the cosmological constant
is at least as small as this.  Similarly, dimensional analysis
suggests that the Higgs mass should be of order $M_p$, but in the
electroweak theory it must be many orders of magnitude smaller.

't Hooft gave a precise statement
of a widely held notion of naturalness\cite{thooftcargese}.  He argued
that if a quantity is much smaller than expected from dimensional
analysis, this should be because the theory becomes more symmetric
in the limit that the quantity tends to zero.  This is a familiar
story for the quark and lepton masses; in the limit that the
corresponding Yukawa couplings tend to zero, the standard model
acquires additional chiral symmetries.  As a result, quantum
corrections to the Yukawa couplings vanish as the couplings tend
to zero.

String theory, on the other hand, sometimes provides exceptions to
this rule, and indeed it does provide an exception in the case of
the cosmological constant.  There are many (non-supersymmetric)
vacua of string theory in which the cosmological constant vanishes
at tree level.  This is already surprising, and a violation of 't
Hooft's rule.  The vanishing of the cosmological constant in these
theories is a consequence not of a symmetry in space-time, but of
a world-sheet symmetry of string theory, {\it conformal
invariance}.  It is an example of what is technically known as a
``string miracle"\cite{miracles}.

On the other hand, for most such states, a cosmological constant
is generated at one loop, and this is far too large to be
compatible with the observed number.  At one loop, in field
theory, the cosmological constant is given by
\beq
\Lambda = \sum_{helicities}(-1)^F{1 \over 2} \int {d^3p \over
(2\pi)^3} \sqrt{\vec p^2 + m_i^2}
\eeq
This expression is correct in (weakly coupled) string theory,
provided that the sum over states is suitably
interpreted\cite{rohm,polchinskicosmo}.  For non-supersymmetric
states, it is typically of order coupling constants times the
string tension to the appropriate power.

In supersymmetric states, $\Lambda=0$, typically, at least in
perturbation theory.  On the other hand, supersymmetry must be
broken in nature, and from the formula above, we expect that
\beq
\Lambda = m_{susy}^4
\eeq
This is compatible with 't Hooft's naturalness principle.  But we
expect that $m_{susy}$ is not smaller than $100~ {\rm GeV}$, so we
are still at least $55$ orders of magnitude off.

This problem is perhaps the most serious obstacle to understanding
string phenomenology.  In these lectures, we will not offer any
solutions.  Only a small number of ideas have been proposed,
and
they are, as yet (at best) incomplete.  Here we mention two:
\begin{itemize}
\item
Witten has noted that in three dimensions, supersymmetry can be
unbroken without degeneracy between bosons and fermions\cite{wittencosmo}.
He
imagines a three dimensional theory in which
there is a single
modulus.  Now suppose that one takes the limit of strong coupling.
Typically, one expects that this is a theory with one additional
dimension.  Perhaps this four dimensional limit is a theory with
broken supersymmetry but vanishing cosmological constant.
Dabholkar and Harvey have constructed models
in various dimensions with small numbers of moduli\cite{dh}.
\item
Kachru and Silverstein, motivated in part by the AdS/CFT correspondence,
have constructed models without supersymmetry in which the
cosmological constant vanishes at low orders of perturbation
theory\cite{ks}.  Conformal invariance plays a crucial role in these
constructions, and the existing examples have Bose-Fermi
degeneracy.  Still, this is perhaps the most successful proposal
to date.
\end{itemize}

Let us turn now to the hierarchy problem.  This is a similar
problem of dimensional analysis.  We might expect that
$m_H^2 \sim M_p^2$, but we
know that if there is a fundamental Higgs scalar, its mass
must be less than about $1$ TeV.  The standard model does not become
more symmetric in the limit that the Higgs mass becomes small,
so this would seem to be a violation of 't Hooft's notion
of naturalness.

String theory again offers an interesting perspective on this
question.  In many weak ground states, at weak
coupling, there
are massless particles, whose masses are not protected by any
symmetry.  This is already a violation of naturalness, and can
usually be understood in terms of world sheet symmetries. In
general, however, one expects radiative corrections to these
masses, just as for the cosmological constant.  For the Higgs
particle in the standard model, for example, loops of gauge bosons
give corrections to the Higgs mass:
\beq
m_H^2 \propto {g^2 \over 16 \pi^2} \int {d^4 k \over (2 \pi)^4 k^2}
\eeq
In string theory, we expect this quadratically divergent integral
will be cut off at the string scale.  Unlike the case of the
cosmological constant, however, for the problem of scalar masses
supersymmetry can offer a resolution.  If the state is
approximately
supersymmetric, then the integrals are cut off, not at the string
scale, but at the scale of supersymmetry breaking, $m_{susy}^2$.
In terms of
Feynman diagrams, one has, due to supersymmetry, more types of
particles, and one finds cancellations\cite{dinetasi}.

If this is the correct explanation, it requires
that $m_{susy}^2 $ not be significantly larger than, say, $1~ {\rm
TeV}$.  So we can hope for discovery soon, perhaps even at LEPII
or at the Tevatron, and certainly at the LHC.

In string theory, these considerations aside, supersymmetry seems
to play an important role, perhaps suggestive of low energy
supersymmetry.  It is easy to find states which, in some
approximation, have low energy supersymmetry ($N=1,2,4,8$ in four
dimensional counting).  We will see, also,  that there are
mechanisms for breaking supersymmetry at low energies.  But we
will need to ask:  to what degree is unbroken supersymmetry a
fundamental property of string theory, and to what degree is it
simply a crutch which gives us some theoretical control?

In any case, supersymmetry is important to our present
understanding of string/M theory.  It provides us with a great
deal of control over dynamics.  It is, for example, the basic of
all of our current understanding of the many dualities of the
theory, as well as of possible non-perturbative formulations (also
dualities) such as matrix theory and the AdS/CFT correspondence.  We
will use this power throughout these lectures.

An alternative proposal for understanding the hierarchy problem
is to suppose that the fundamental scale is not $M_p$ but
actually of order $1$ TeV.  This requires that there be some
large internal dimensions\cite{largeradii,precursors},
or a suitable ``warp factor" in the
extra dimensions\cite{rs}.  These possibilities have received much
attention in the last year.  At this point, they do not appear
significantly less plausible than low energy supersymmetry as a
solution.  We will focus in these lectures mostly on
supersymmetry, since, as we will see, it is, within our current
understanding, easier to develop
scenarios for the realization of supersymmetry in string theory
where there is some understanding of why couplings are weak and of
which quantities might be calculable.   But given our lack
of a complete picture, this may well reflect simply ``the state of
the art."   We will make some comments on these ideas,
particularly in section 9.

The rest of these lectures will be devoted to developing tools for
thinking about string dynamics and string phenomenology, based
largely on supersymmetry.  The next section presents a brief
overview of supersymmetry, its representations, the structure of
supersymmetric lagrangians (global and local), and the use of
superspace.  Section 3 discusses some quantum aspects of
supersymmetric theories.  For theories with $N=2$ and $N=4$
supersymmetry, we will see that moduli are exact quantum
mechanically, and prove certain non-renormalization theorems.  For
the case of $N=1$ supersymmetry, we will see that fields which are
moduli in some approximation generically have non-trivial
potentials in the full theory.  We will
introduce the notion of {\it approximate moduli}, discuss
supersymmetry breaking, and also consider circumstances under
which moduli are exact.
The fourth section consists of a brief review of $N=1$
supersymmetry phenomenology:  the Minimal Supersymmetric Standard
Model (MSSM), soft breakings, counting parameters, constraints,
direct detection and theories of soft breakings.

After this, we turn to string theory.  Section 5 focuses on string
moduli.
It is possible to make many exact statements about the full,
microscopic theory by focusing on a low energy effective
lagrangian for the light fields.
We explain why, in much the same way as for field theory, one can discuss
the exactness of string moduli and issues of non-renormalization.
Even in theories with $N=1$ supersymmetry, in some cases the
problem of supersymmetry breakdown is a problem of low energy
physics.   We discuss the issue of modulus stabilization, and
moduli in cosmology.  We make some tentative statements about
string phenomenology.

The sixth section is devoted to string phenomenology in the light
of duality.  We will focus particularly on the Horava-Witten
picture, the role of branes, and on recent proposals that the
string scale might be as low as $1 ~{\rm TeV}$.

In the seventh section we will asses the outlook for achieving the
goal we  set forth in the beginning, of obtaining a few robust
predictions from string theory.  Overall, we will cover many
topics, and offer some  speculations, but we won't provide any
real answers to the big questions.

\section{An Overview of Supersymmetry}

\subsection{The Supersymmetry Algebra and its Representations}

In this lecture, we will collect a few facts that will be useful
in the subsequent discussion.  We won't attempt a thorough
introduction to the subject.  This is provided, for example, by
Lykken's lectures\cite{lykken,superspace}, Wess and Bagger's text\cite{wb},
and Appendix B of Polchinski's text\cite{polchinski}.

Supersymmetry, even at the global level, is remarkable, in that
the basic algebra involves the translation generators: \beq
\{Q_{\alpha}^A,Q^{*B}_{\dot \beta}\}=2\sigma^{\mu}_{\alpha \dot
\beta}\delta^{AB} P_{\mu}
\label{susyalgebra}
\eeq
\beq
\{Q_{\alpha}^A,Q^*_{\beta B}\}= \epsilon_{\alpha \beta}X^{AB}.
\eeq
The $X^{AB}$'s are Lorentz scalars, antisymmetric in $A,B$, known
as central charges.

If nature is supersymmetric, it is likely that the
low energy symmetry is $N=1$, corresponding to only one possible
value for the index $A$ above.  Only $N=1$ supersymmetry has
chiral representations.  In addition,
$N>1$ supersymmetry, as we will see, is essentially impossible to
break; this is not the case for $N=1$.   For $N=1$, the basic
representations of the supersymmetry algebra, on massless fields, are
\begin{itemize}
\item
Chiral superfields fields:  $(\phi,\psi_{\alpha})$, a complex fermion and a
chiral scalar
\item
Vector superfields:
$(\lambda,A_{\mu})$, a chiral fermion and a vector meson, both, in
general, in the adjoint representation of the gauge group
\item
The gravity supermultiplet:
$(\psi_{\mu,\alpha},g_{\mu\nu})$,
a spin-3/2 particle, the gravitino, and the graviton.
\end{itemize}

N=1 supersymmetric field theories are conveniently described using
superspace.  The space consists of bosonic coordinates, $x^{\mu}$,
and Grassman coordinates, $\theta_{\alpha},\theta_{\dot \alpha}^*$.
In the case of global supersymmetry, the description is particularly
simple.  The
supersymmetry generators, classically, can be thought of as
operators on functions of $x^{\mu}$, $\theta,\theta^*$:
\beq
Q_{\alpha}= \partial_{\alpha}-
i\sigma^{\mu}_{\alpha \dot \alpha} \theta^{*\dot
\beta}\partial_{\mu};
~~~~~\bar Q_{\dot \alpha}= -\partial_{\dot \alpha}+
i
\theta^{*\alpha}\sigma^{\mu}_{\alpha \dot \alpha}\partial_{\mu}.
\eeq

A general superfield, $\Phi(x,\theta,\bar \theta)$ contains many
terms, but can be decomposed into two irreducible representations
of the algebra, corresponding to the chiral and vector superfields
described above.  To understand these, we need to introduce one
more set of objects, the covariant derivatives, $D_{\alpha}$
and $\bar D_{\dot \alpha}$.  These are objects which anti-commute with the
supersymmetry generators, and thus are useful for writing down
invariant expressions.  They are given by
\beq
D_{\alpha}= \partial_{\alpha}
i\sigma^{\mu}_{\alpha \dot \alpha} \theta^{* \dot
\alpha}\partial_{\mu};
~~~~~\bar D_{\dot \alpha}= -\partial_{\dot \alpha}-
i
\theta^{\alpha}\sigma^{\mu}_{\alpha \dot \alpha}\partial_{\mu}.
\eeq

With this definition, chiral fields are defined by the covariant
condition:
\beq
\bar D_{\dot \alpha}\Phi=0.
\label{chiralcondition}
\eeq

Chiral fields are annihilated by the covariant
derivative operators.  In general, these covariant derivatives
anticommute with the supersymmetry operators, $Q_{\alpha}$, so the
condition \ref{chiralcondition}
is a covariant condition.  This is solved by writing
\beq
\Phi = \Phi(y) = \phi(y) + \sqrt{2} \theta \psi(y) + \theta^2
F(y).
\eeq
where
\beq
y=x^{\mu} + i \theta \sigma^{\mu} \bar \theta.
\eeq

Vector superfields form another irreducible representation of the
algebra; they satisfy the condition
\beq
 V= V^{\dagger}
\eeq
Again, it is easy to check that this condition is preserved by
supersymmetry transformations.  $V$ can be expanded in a power
series in $\theta$'s:
\beq
V= i\chi -i \chi^{\dagger} - \theta \sigma^{\mu} \theta^* A_{\mu}
+i \theta^2 \bar \theta \bar \lambda - i \bar \theta^2 \theta
\lambda + {1 \over 2} \theta^2 \bar \theta^2 D.
\eeq
In the case of a $U(1)$ theory,
gauge transformations act by
\beq
V \rightarrow V+ i\Lambda -i \Lambda^{\dagger}
\eeq
where $\Lambda$ is a chiral field.  So, by a gauge transformation,
one can eliminate $\chi$.  This gauge choice is known as the
Wess-Zumino gauge.  This gauge choice breaks supersymmetry, much
as choice of Coulomb gauge in electrodynamics breaks Lorentz
invariance.

In the case of a $U(1)$ theory, one can define a gauge-invariant
field strength,
\beq
W_{\alpha} = -{1 \over 4} \bar D^2 D_{\alpha} V.
\eeq
In Wess-Zumino gauge, this takes the form
\beq
W_{\alpha}=
-i \lambda_{\alpha} + \theta_{\alpha} D
-{\sigma^{\mu \nu}}_{\alpha}^{\beta}F_{\mu \nu}\theta_{\beta}
+\theta^2 \sigma^{\mu}_{\alpha \dot \beta}\partial_{\mu}
\lambda^{* \dot \beta}.
\eeq
This construction has a straightforward non-Abelian
generalization in superspace, which is described in the references.
When we write the lagrangian in terms of component fields below,
the non-abelian generalization will be obvious.

\subsection{N=1 Lagrangians}

One can construct invariant lagrangians by noting that integrals
over superspace are invariant up to total derivatives:
\beq
\delta \int d^4 x \int d^4 \theta ~h(\Phi,\Phi^{\dagger},V) =
\int d^4 x d^4 \theta~(\epsilon_{\alpha} Q^{\alpha}+ \epsilon_{\dot \alpha} Q^{\dot
\alpha}) h(\Phi,\Phi^{\dagger},V) = 0.
\eeq
For chiral fields, integrals over {\it half} of superspace are
invariant:
\beq
\delta \int d^4 x d^2 \theta f(\Phi) =
(\epsilon_{\alpha} Q^{\alpha}+ \epsilon_{\dot \alpha} Q^{\dot
\alpha})  f(\Phi).
\eeq
The integrals over the $Q_{\alpha}$ terms vanish when integrated
over $x$ and $d^2 \theta$.  The $Q^*$ terms also give zero.  To
see this, note that
$f(\Phi)$ is itself chiral (check), so
\beq
 Q^*_{\dot \alpha} f
\propto \theta^{\alpha} {\sigma^{\mu}}_{\alpha \dot
\alpha} \partial_{\mu} f.
\eeq

We can then write down the general renormalizable, supersymmetric
lagrangian:
\beq
{\cal L} = {1 \over g^{(i)2}}\int d^2 \theta W_{\alpha}^{(i) 2}
+ \int d^4 \theta \Phi_i^{\dagger} e^{q_i V} \Phi_i \int d^2 \theta W(\Phi_i)
+ {\rm c.c.}
\eeq
The first term on the right hand side is summed over all of the
gauge groups, abelian and non-abelian.  The second term is summed
over all of the chiral fields; again, we have written this for a
$U(1)$ theory, where the gauge group acts on the $\Phi_i$'s by
\beq
\Phi_i \rightarrow  e^{-q_i \Lambda} \Phi_i
\eeq
but this has a simple non-abelian generalization.  $W(\Phi)$ is a
holomorphic function of the $\Phi_i$'s (it is a function of
$\Phi_i$, not $\Phi_i^{\dagger}$).

In terms of component fields, his lagrangian takes the form, in
the Wess-Zumino gauge:
\beq {\cal L} = -{1 \over 4} g_a^{-2} F_{\mu \nu}^{a2}
-i\lambda^a \sigma^{\mu}D_{\mu}
 \lambda^{a*} +
+\vert D_{\mu}\phi_i \vert^2 -i \psi_i \sigma^{\mu}D_{\mu}
\psi_i^* + {1 \over 2 g^2} (D^a)^2 + D^a\sum_i \phi_i^* T^a
\phi_i
\eeq
$$~~~~~~~+ F_i^*F_i -F_i {\partial W \over \partial \phi_i}
+{\rm cc} + \sum_{ij}{1 \over 2}{\partial^2 W \over \partial
\phi_i
\partial \phi_j}\psi_i \psi_j + i \sqrt{2}\sum \lambda^a \psi_i T^a
\phi_i^*.$$

The scalar potential is found by solving for the auxiliary $D$ and
$F$ fields:
\beq
V= \vert F_i \vert^2 +{1 \over 2 g_a^2} (D^a)^2
\eeq
with
\beq
F_i = {\partial W \over \partial \phi_i^*}
~~~~~~~~~D^a = \sum_i(g^a\phi_i^* T^a \phi_i).
\eeq
In this equation, $V \ge 0$.  This fact can be traced back
to the supersymmetry algebra.  Starting with the equation
\beq
\{Q_{\alpha},Q_{\dot \beta}\} = 2 P_{\mu}\sigma^{\mu}_{\alpha \dot
\beta},
\eeq
Multiplying  by $\sigma^o$ and take the trace:
\beq
Q_{\alpha}Q_{\dot \alpha} + Q_{\dot \alpha} Q_{\alpha} = E.
\eeq
If supersymmetry is unbroken,
$Q_{\alpha}\vert 0 \rangle =0$, so
the ground state energy vanishes {\it if and only if}
supersymmetry is unbroken.
Alternatively, consider the
supersymmetry transformation laws for $\lambda$ and $\psi$.  One
has, under a supersymmetry transformation with parameter
$\epsilon$,
\beq
\delta \psi = \sqrt{2}\epsilon F + \dots~~~~~~\delta \lambda = i\epsilon D
+ \dots
\eeq
So if either $F$ or $D$ has an expectation value, supersymmetry is
broken.

We should stress that these statements apply to global
supersymmetry.  We will discuss the case of local supersymmetry
later, but, as we will see, many of the lessons from the global
case extend in a simple way to the case in which the symmetry is a
gauge symmetry.

We can now very easily construct a supersymmetric version of the standard
model.  For each of the gauge fields of the usual standard model,
we introduce a vector superfield.  For each of the fermions
(quarks and leptons) we introduce a chiral superfield with the
same gauge quantum numbers.  Finally, we need at least two Higgs
doublet chiral fields; if we introduce only one, as in the
simplest version of the standard model, the resulting theory
possesses gauge anomalies and is inconsistent.  In other words,
the theory is specified by giving the gauge group ($SU(3)\times
SU(2) \times U(1)$)
and enumerating the chiral fields:
\beq
Q_f, \bar u_f, \bar d_f ~~~~~
L_f, \bar e_f
~~~~~
H_U, H_D.\eeq

The gauge invariant kinetic terms, auxiliary $D$ terms, and
gaugino-matter Yukawa couplings are completely specified by the
gauge symmetries.  The superpotential can be taken to be:
\beq
W=H_U (\Gamma_{U})_{f,f^{\prime}}Q_f \bar U_{f^{\prime}}
+H_D(\Gamma_{D})_{f,f^{\prime}}Q_f \bar D_{f^{\prime}}
H_D(\Gamma_{E})_{f,f^{\prime}}L_f \bar e_{f^{\prime}}
.
\eeq
As we will discuss shortly, this is not the most general
lagrangian consistent with the gauge symmetries.  It does yield
the desired quark and lepton mass matrices, without other
disastrous consequences.

\noindent
{\bf Exercise}:  Consider the case of one generation.  Show that
if
\beq
\langle H_U \rangle = \langle H_D \rangle = \left ( \matrix{0 \cr
v} \right ),
\eeq
(all others vanishing), then
\beq
\langle D^a \rangle =0; \langle F^i \rangle =0.
\eeq
Study the spectrum of the model.  Show that the superpartners of
the $W$ and $Z$ are degenerate with the corresponding gauge
bosons.  (Note that for the massive gauge bosons, the multiplet
includes an additional scalar).  Show that the quarks and leptons
gain mass, and are degenerate with their scalar partners.

The fact that the states fall into degenerate multiplets reflects
that for this set of ground states (parameterized by $v$),
supersymmetry is unbroken.  That supersymmetry is unbroken follows
from the fact that the energy is zero, by our earlier argument.
It can also be understood by examining the transformation laws for
the fields.  For
example,
\beq
\delta \phi_i = \zeta_{\alpha}[Q^\alpha,\phi_i] = \sqrt{2} i \zeta
\psi
\eeq
but the right hand side has no expectation value.  Similarly,
\beq
\delta \psi_i = \sqrt{2} \zeta F_i + \sqrt{2}
i \sigma^{\mu} \bar \zeta \partial_{\mu} \phi
.
\eeq
The last term vanishes by virtue of the homogeneity of the ground
state; the first vanishes because $F_i=0$.  Similar statements
hold for the other possible transformations.

This is our first example of a moduli space.  Classically, at
least, the energy is zero for any value of $v$.  So we have a one
parameter family of ground states.  These states are physically
inequivalent, since, for example, the mass of the gauge bosons
depends on $v$.  We will shortly explain why, in field theory, it
is necessary to choose a particular $v$, and why there are not
transitions between states of different $v$ (in any approximation
in which degeneracy holds).   As we will see later in these
lectures, generically classical moduli spaces are not moduli
spaces at the quantum level.

\begin{figure}[htbp]
\centering
\centerline{\psfig{file=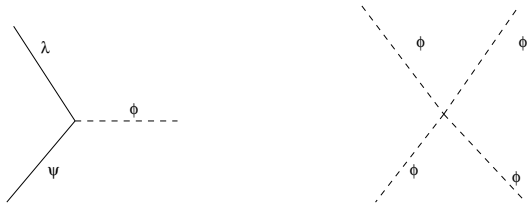,angle=-90,width=7cm}}
\caption{Some of the vertices in a supersymmetric theory.
Dashed lines denote scalars, solid lines fermions.}
\label{somecouplings}
\end{figure}

One can also read off from the lagrangian the couplings, not only
of ordinary fields, but of their superpartners.  For example,
there is a Yukawa coupling of the gauginos to fermions and
scalars, whose strength is governed by the corresponding gauge
couplings.  There are also quartic couplings of the scalars, with
gauge strength.  These are indicated in fig. \ref{somecouplings}.

Before turning to the phenomenology of this ``Minimal
Supersymmetric Standard Model," (MSSM), it is useful to get some
more experience with the properties of supersymmetric theories.
With the limited things we know, we can already derive some
dramatic results.  First, we can write down the most general
globally supersymmetric lagrangian, with terms with at most two
derivatives, but not restricted by renormalizability (in the rest
of this section, the lower case $\phi$ refers both to the
chiral field and its scalar component):
\beq
{\cal L}= \int d^4 \theta K(\phi_i^{\dagger},\phi_i)
+ \int d^2 \theta W(\phi_i) + {\rm cc}
+ \int d^2 \theta f(\phi_i)W_{\alpha}^2 + {\rm cc}.
\eeq
Here $K$ is a general function known as the Kahler potential.  $W$
and $f$ are necessarily {\it holomorphic} functions of the chiral
fields.  One can consider terms involving the covariant
derivatives, $D_{\alpha}$, but these correspond to terms with more
than two derivatives, when written in terms of component fields.

We will often be interested in effective lagrangians of
this sort, for example, in studying the low
energy limit of string theory.  From the holomorphy of $W$ and
$f$, as well as from the symmetries of the models, one can often
derive remarkable results.  Consider, for example, the
``Wess-Zumino" model, a model with a single chiral field with
superpotential
\beq
W= {1 \over 2} m \phi^2 + {1 \over 3} \lambda \phi^3.
\eeq
For general $m$ and $\lambda$, this model has no continuous global
symmetries.  If $m=0$, is has an ``$R$" symmetry, a symmetry which
does not commute with supersymmetry:
\beq
\phi \rightarrow e^{{2 i\over 3}\alpha}\phi
~~~~~\theta \rightarrow e^{i \alpha }\theta
~~~~~~d \theta \rightarrow e^{-i \alpha} \theta.
\eeq
Under this transformation,
\beq
W \rightarrow e^{2 i \alpha}W
\eeq
so $\int d^2 \theta W$ is invariant.
This transformation does not commute with supersymmetry; recalling
the form of $Q_{\alpha}$ in terms of $\theta$'s, one sees that
\beq
Q_{\alpha} \sim {\partial \over \partial \theta_{\alpha} } +
\dots \rightarrow e^{i \alpha} Q_{\alpha}.
\eeq
Correspondingly, the fermions and scalars in the multiplet transform
differently: the scalar has the same $R$ charge as the
superfield, $2/3$,
while $\psi$ has R charge one unit less than that of the scalar,
i.e.
\beq
\phi \rightarrow e^{2 i \alpha/3} \phi ~~~~~~~\psi \rightarrow
e^{-i \alpha/3} \psi.
\eeq
It is easy to check that this a symmetry of the lagrangian,
written
in terms of the component fields.
Correspondingly, in the quantum theory,
\beq
Q_{\alpha} \approx \int d^3 x (\sigma^{\mu}_{\alpha \dot \alpha}
\partial_{\mu}\phi \psi^{*\dot \alpha} +
\psi_{\alpha}F) \rightarrow e^{i\alpha}Q_{\alpha}.
\eeq

Symmetries of this type will play an important role in much of
what follows.  In general, in a theory with several chiral fields,
one has
\beq
\phi_i \rightarrow e^{i \alpha R_i} \phi_i~~~~~~~W(\phi_i) \rightarrow
e^{2 i \alpha} W(\phi_i).
\eeq
If there are vector multiplets in the model, the gauge bosons are
neutral under the symmetry, while the gauginos have charge $+1$.
We will also be interested in discrete versions of these
symmetries (in which, essentially, the parameter $\alpha$ takes on
only some discrete values).
In the case of the Wess-Zumino model,  for non-zero $m$, a
discrete subgroup survives for which $\alpha= 3n\pi$, i.e.
$\phi \rightarrow \phi$, $\psi \rightarrow - \psi$.  More
elaborate discrete symmetries will play an important role in our
discussions.

Even in the Wess-Zumino model with non-zero $m$, we can exploit
the power of continuous symmetries, by thinking of the couplings as
if they were themselves background values for some
chiral fields.  If we assign $R$ charge $+1$
to $\phi$, we can make the theory $R$-invariant if we assign $R$
charge $-1$ to $\lambda$.  In practice, we might be interested in
theories where $\lambda$ is the scalar component of a dynamical
field (this will often be the case in string theory) or we may
simply view this as a trick\cite{seibergnr}.  In either case, we
can immediately see that there are no corrections to the $\phi^3$
term in the superpotential in powers of $\lambda$.  The reason is
that the superpotential must be a holomorphic function of $\phi$
{\it and} $\lambda$ (so it cannot involve,
say, $\lambda \lambda^{\dagger}$), and it must respect the $R$ symmetry.
Because $\lambda$ is the small parameter of the theory, we
have proven a powerful non-renormalization theorem:  the
superpotential cannot be corrected to any order in the coupling
constant.
This  non-renormalization theorem was
originally derived by detailed consideration of the properties of
Feynman graphs.  What is crucial to this argument is that $W$ is a
holomorphic function of $\phi$ {\it and} the parameters of the
lagrangian.

This is not to say that nothing in the effective
action of the theory is corrected
from its lowest order value; non-holomorphic quantities
are renormalized. For example:
\beq
\int d^4 \theta \phi^{\dagger}\phi f(\lambda^{\dagger} \lambda)
\eeq
is allowed.  In the Wess-Zumino model, this means that all of the
renormalizations are determined by wave function renormalization.
Finally, we should note that if $m=0$ at tree level, no masses are
generated for fermions or scalars in loops.

\subsection{N=2 Theories:  Exact Moduli Spaces}

We have already encountered an extensive vacuum degeneracy in the
case of the MSSM.  Actually, the degeneracy is much larger;
there is a multiparameter family of such flat directions
involving the squark, slepton and Higgs fields.  For the
particular example,
we saw that classically the possible ground
states of the theory are labeled by a quantity $v$.  States with
different $v$ are physically distinct; the masses of particles,
for example, depend on $v$.  In non-supersymmetric theories, one doesn't
usually contemplate such degeneracies, and even if one had such a degeneracy,
say, at the classical level, one would expect it to be
eliminated by quantum effects.  We will see that in supersymmetric
theories, these flat directions
almost always remain flat in perturbation theory; non-perturbatively,
they are sometimes lifted, sometimes not.  Moreover, such directions are
ubiquitous  The space of degenerate ground
states of a theory is called the ``moduli space."  The fields whose
expectation values label these states are called the
moduli. In supersymmetric theories, such
degeneracies are common, and are often not spoiled by quantum
corrections.

In theories with $N=1$ supersymmetry, detailed
analysis is usually required to determine whether the moduli
acquire potentials at the quantum level.  For theories with more
supersymmetries ($N>1$ in four dimensions; $N \ge 1$ in five or
more dimensions), one can usually show rather easily that the
moduli space is exact.  Here we consider the case of $N=2$
supersymmetry in four dimensions.  These theories can also be
described by a superspace, in this case built from two Grassman
spinors, $\theta$ and $\tilde \theta$.  There are two basic types
of superfields\cite{lykken}, called vector and hyper multiplets.
The vectors are chiral with respect to both $D_{\alpha}$ and
$\tilde D_{\alpha}$, and have an expansion, in the case of a $U(1)$
field:
\beq
\psi = \phi + \tilde \theta^{\alpha} W_{\alpha}
+ \tilde \theta^2 \bar D^2 \phi^{\dagger},
\eeq
where $\phi$ is an $N=1$ chiral multiplet and $W_{\alpha}$
is an $N=1$ vector multiplet.
The fact that $\phi^{\dagger}$ appears as the coefficient of the
$\tilde \theta^2$ term is related to an additional constraint
satisfied by $\psi$\cite{lykken}.  This expression can be
generalized to non-abelian symmetries; the expression for the
highest component of $\psi$ is then somewhat more complicated\cite{lykken}; we
won't need this here.

The theory possesses an $SU(2)$ R symmetry under which $\theta$
and $\tilde \theta$ form a doublet.  Under this symmetry, the
scalar component of $\phi$, and the gauge field, are singlets,
while $\psi$ and $\lambda$ form a doublet.

I won't describe the hypermultiplets in detail, except to note
that from the perspective of $N=1$, they consist of two chiral
multiplets.  The two chiral multiplets transform as a
doublet of the $SU(2)$.  The superspace description of these
multiplets is more complicated\cite{lykken,superspace}.

In the case of a non-abelian theory, the vector field, $\psi^a$,
is in the adjoint representation of the gauge group.
For these fields, the lagrangian has a very simple expression
in superspace:
\beq
{\cal L}= \int d^2 \theta d^2 \tilde \theta~ \psi^a \psi^a
,\eeq
or, in terms of $N=1$ components,
\beq
{\cal L}=\int d^2 \theta~ W_{\alpha}^2 + \int d^4 \theta
\phi^{\dagger} e^V \phi.
\eeq
The theory with vector fields alone has a
classical moduli space, given by the values of the fields for
which the scalar potential vanishes. Here this just means that the $D$
fields vanish.  Written as a matrix,
\beq
D = [\phi,\phi^{\dagger}],
\eeq
which
vanishes for diagonal $\phi$, i.e. for
\beq
\phi = {a \over 2} \left ( \matrix{1 & 0 \cr 0 & -1} \right )
.
\eeq

In quantum field theory, one must choose a value of $a$.   This
is different than in the case of quantum mechanical systems
with a finite number of degrees of freedom; this difference
will be explained below.  As in
the case of the MSSM, the spectrum depends on $a$.  For a given
value of $a$, the massless states consist of a $U(1)$ gauge boson,
two fermions, and a complex scalar (essentially $a$), i.e. there
is one light vector multiplet.  The masses of the states in the
massive multiplets depend on $a$.

For many physically interesting questions, one can focus on the
effective theory for the light fields.  In the present case, the light
field is the vector multiplet, $\psi$.  Roughly,
\beq
\psi \approx \psi^a \psi^a = a^2 + a \delta \psi^3 + \dots
\eeq
What kind of effective action can we write for
$\psi$?
At the level of terms with up to four derivatives, the most
general effective lagrangian has the form:\footnote{This, and
essentially all of the effective
actions we will discuss, should be thought of as Wilsonian
effective actions, obtained by integrating out heavy fields and
high momentum modes.}
\beq
{\cal L}= \int d^2 \theta d^2 \tilde \theta f(\psi)
+ \int d^8 \theta {\cal H}(\psi,\psi^{\dagger}).
\label{nequalstwolagrangian}
\eeq
Terms with covariant derivatives correspond to terms with more
than four derivatives, when written in terms of ordinary component
fields.

The first striking result we can read off from this lagrangian,
with no knowledge of ${\cal H}$ and $f$, is that there is no
potential for $\phi$, i.e. the moduli space is exact.  This
statement is true perturbatively and non-perturbatively!

One can next ask about the function $f$.  This function determines
the effective coupling in the low energy theory, and is the object
studied by Seiberg and Witten\cite{seibergwitten}.
We won't review this whole story here, but indicate how symmetries
and the holomorphy of $f$ provide significant constraints (Michael
Peskin's TASI 96 lectures provide a concise introduction to this
topic\cite{peskintasi}).  It is helpful, first, to introduce a
background field, $\tau$, which we will refer to as the
``dilaton," with coupling
\beq
{\cal L}= \int d^2 \theta d^2 \tilde \theta \tau \psi^a \psi^a
\eeq
where
\beq
\tau = \theta + {i \over g^2} + \dots.
\eeq
$\tau$ is a chiral field.  For our purposes, $\tau$ need not be
subject to the same constraint as the vector superfield.
Classically, the theory has an R-symmetry
under which $\psi^a$ rotates by a phase,
$\psi^a \rightarrow  e^{i \alpha} \psi^a$.
But this symmetry is anomalous.  Similarly, shifts in the real
part of $\tau$ ($\theta$) are symmetries of perturbation theory.
This insures that there is only a one-loop correction to $f$.
This follows, first, from the fact that any perturbative
corrections to $f$
must be $\tau$-independent.  A term of the form
\beq
c \ln(a) \psi^2
\eeq
respects the symmetry, since the shift of the logarithm just
generates a contribution proportional to $F \tilde F$, which
vanishes in perturbation theory.  Beyond perturbation theory,
however, we expect corrections proportional to $a e^{-\tau}$,
since this is invariant under the non-anomalous symmetry.  It is
these corrections which were worked out by Seiberg and Witten.

\subsection{A Still Simpler Theory:  N=4 Yang Mills}

$N=4$ Yang Mills theory is an interesting theory in its own right:
it is finite and conformally invariant.  It also plays an
important role in Matrix theory, and is central to our
understanding of the AdS/CFT correspondence.
$N=4$ Yang Mills has sixteen supercharges, and is even more
tightly constrained than the $N=2$ theories.
There does not exist a convenient superspace
formulation for this theory, so we will find it necessary to resort to
various tricks.  First, we should describe the theory.  In the
language of $N=2$ supersymmetry, it consists of one vector and one
hyper multiplet.  In terms of $N=1$ superfields, it contains three
chiral superfields, $\phi_i$, and a vector multiplet.  The
lagrangian is
\beq
{\cal L}= \int d^2 \theta W_{\alpha}^2 + \int d^4 \theta
\sum\phi_i^{\dagger} e^V \phi_i + \int d^2 \theta \phi_i^a
\phi_j^b \phi_k^c \epsilon_{ijk}\epsilon^{abc}.
\eeq
In the above description, there is a manifest $SU(3) \times U(1)$
R symmetry.  Under this symmetry, the $\phi_i$'s have $U(1)_R$
charge $2/3$, and form a triplet of the $SU(3)$.  But the real
symmetry is larger -- it is $SU(4)$.  Under this symmetry, the
four Weyl fermions form a $\bf 4$, while the six scalars transform in
the $6$.  Thinking of these theories as the low energy limits of
toroidal compactifications of the heterotic string will later give
us a heuristic understanding of this $SU(4)$:  it reflects the
$O(6)$ symmetry of the compactified six dimensions.  In string theory,
this symmetry
is broken by the compactification lattice; this reflects itself in
higher derivative, symmetry breaking operators.

In the $N=4$ theory, there is, again, no modification of the
moduli space, perturbatively or non-perturbatively.  This can be
understood in a variety of ways.  We can use the $N=2$
description of the theory, defining the vector multiplet to
contain the $N=1$ vector and one (arbitrarily chosen) chiral
multiplet.  Then an identical argument to that given above insures
that there is no superpotential for the chiral multiplet alone.
The $SU(3)$ symmetry then insures that there is no superpotential
for any of the chiral multiplets.  Indeed, we can make an argument
directly in the language of $N=1$ supersymmetry.  If we try to
construct a superpotential for the low energy theory in the flat
directions, it must be an $SU(3)$-invariant, holomorphic function
of the $\phi_i$'s.  But there is no such object.

Similarly, it is easy to see that there no corrections to the
gauge couplings.  For example, in the $N=2$ language, we want to
ask what sort of function, $f$, is allowed in
\beq
{\cal L}= \int d^2 \theta d^2 \tilde \theta f(\psi).
\eeq
But the theory has a $U(1)$ R invariance under which
\beq
\psi \rightarrow e^{2/3 i \alpha} \psi~~~~~~~~\theta \rightarrow
e^{i\alpha \theta} ~~~~~~\tilde \theta \rightarrow
e^{-i\alpha}\tilde \theta
\eeq
Already, then
\beq
\int d^2 \theta d^2 \tilde \theta \psi \psi
\eeq
is the unique structure which respects these symmetries.
Now we can introduce a background dilaton field, $\tau$.
Classically, the theory is invariant under shifts in the real
part of $\tau$, $\tau \rightarrow \tau +  \beta$.  This insures
that there are no perturbative corrections to the gauge couplings.
More work is required to show that there are no non-perturbative
corrections either.

One can also show that the quantity ${\cal H}$ in
eqn. [\ref{nequalstwolagrangian}]
is unique in this
theory, again using the symmetries.  The expression\cite{roceketal,dsnr}:
\beq
{\cal H}= c \ln(\psi) \ln(\psi^{\dagger}),
\eeq
respects all of the symmetries.  At first sight, it might
appear to violate scale invariance; given that $\psi$
is dimensionful, one would expect a scale, $\Lambda$, sitting in
the logarithm.  However, it is easy to see that one integrates
over the full superspace, any $\Lambda$-dependence disappears,
since $\psi$ is chiral.  Similarly, if one considers the $U(1)$
R-transformation, the shift in the lagrangian vanishes after the
integration over superspace.  To see that this expression is not
renormalized, one merely
needs to note that any non-trivial $\tau$-dependence spoils
these two properties.  As a result, in the case of $SU(2)$, the
four derivative terms in the lagrangian are not renormalized.
Note that this argument is non-perturbative.

\subsection{Aside:  Choosing a Vacuum}

It is natural to ask:  why in field theory, in the presence of
moduli, does one have to choose a vacuum?  In other words, why
aren't their transition between states with different expectation
values for the moduli?

\begin{figure}[htbp]
\centering
\centerline{\psfig{file=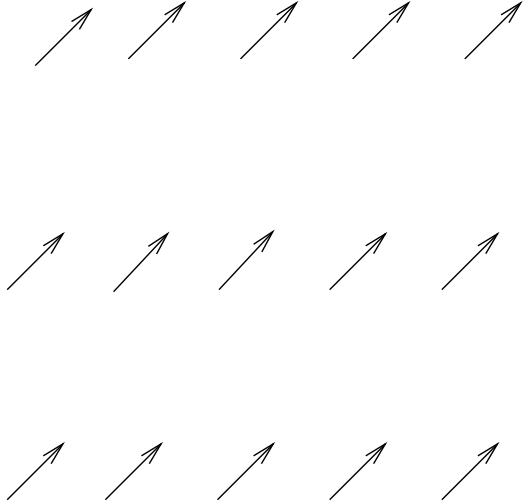,angle=-90,width=7cm}}
\caption{In a ferromagnetic, the spins are aligned, but the direction
is arbitrary.}
\label{magnet}
\end{figure}

This issue is most easily understood by considering a
different problem:  rotational invariance in a magnet.  Consider
fig. \ref{magnet}, where we have considered a ferromagnet with spins
aligned in two different directions, one oriented at an angle
$\theta$ relative to the other.  We can ask:  what is the overlap
of the two states, i.e. what is $\langle \theta \vert 0 \rangle$?
For a single site, the overlap between the state $\vert + \rangle$
and the rotated state is:
\beq
\langle + \vert e^{i \tau_1 \theta/2} \vert + \rangle =
\cos(\theta/2).
\eeq
If there are $N$ such sites, the overlap behaves as
\beq
\langle \theta \vert 0 \rangle \sim (\cos(\theta/2))^N
\eeq
i.e. it vanishes exponentially rapidly with the volume.

For a continuum field theory, states with differing values of the
order parameter, $v$, also have no overlap in the infinite
volume limit.  This is illustrated by the theory of a
scalar field with lagrangian:
\beq
{\cal L}= {1 \over 2} (\partial_{\mu}\phi)^2.
\eeq
For this system, the expectation value $\phi=v$ is
not fixed.  The lagrangian has a symmetry, $\phi \rightarrow \phi
+ \delta$, for which the charge is just
\beq
Q = \int d^3 x \Pi(\vec x)
\eeq
where $\Pi$ is the canonical momentum.  So we want to study
\beq
\langle v \vert 0 \rangle = \langle 0 \vert e^{iQ} \vert 0 \rangle.
\eeq
We must be careful how we
take the infinite volume limit.  We will insist that this be done
in a smooth fashion, so we will define:
\beq
Q= \int d^3 x \partial_o \phi e^{-\vec x^2 /V^{2/3}}
\eeq
$$=~~~~~~~-i \int {d^3 k \over (2 \pi)^3}\sqrt{\omega_k \over 2}
\left ({V^{1/3} \over \sqrt{\pi}} \right )^3 e^{-\vec k^2 V^{2/3}/4}
[a(\vec k) - a^{\dagger}(\vec k)].
$$
Now, one can evaluate the matrix element, using
$$e^{A+B} = e^A e^Be^{{1 \over 2}[A,B]}$$
(provided that the commutator is a c-number), giving
\beq
 \langle 0\vert e^{iQ} \vert 0 \rangle = e^{-c v^2 V^{2/3}},
 \eeq
 where $c$ is a numerical constant.
 So the overlap vanishes with the volume.  You can convince
 yourself that the same holds for matrix elements of local
 operators.
 This result does not hold in $0+1$ and $1+1$
 dimensions, because of the severe infrared behavior of
 theories in low dimensions.  This is known to particle physicists
 as Coleman's theorem, and to condensed matter theorists as the
 Mermin-Wagner theorem.

\section{N=1:  Supersymmetry Breaking?}

In four dimensions, we have seen that in theories with more than
two supersymmetries, moduli are exact and supersymmetry remains
unbroken exactly.  We turn, now, to $N=1$ theories.  We will prove that
if supersymmetry is unbroken at tree level, it
remains unbroken to all orders in perturbation theory.  Non-perturbatively,
however, we will see that supersymmetry is often broken; supersymmetry
breaking is typically of order
$e^{-c 8 \pi^2/g^2}$, where $g$ is some gauge coupling,
and $c$ is a numerical constant.  The potential to generate
large hierarchies under such circumstances was first stressed
by Witten\cite{wittendsb}.

If supersymmetry is spontaneously broken, there is
necessarily a massless fermion, just as breaking of an ordinary
global symmetry implies the existence of a Goldstone boson.  The
proof closely parallels the usual proof of Goldstone's theorem,
and the corresponding massless fermion is called the Goldstino.
For example, for chiral fields, under a supersymmetry
transformation with parameter $\epsilon$,
\beq
\delta \psi = \epsilon F + \dots
\eeq
So if $\langle F \rangle \ne 0$, supersymmetry is broken.  In
terms of the supersymmetry current:
\beq
j^{\mu}_{\alpha} = F\sigma^{\mu}_{\alpha \dot\alpha}\psi^{*\dot
\alpha}+ \dots
\eeq
$$~~~~~~~~\approx \langle F \rangle \sigma^{\mu}_{\alpha \dot
\alpha}\psi^{*\dot \alpha}.$$

Since
\beq
\partial_{\mu}j^{\mu}_{\alpha}=0
\eeq
we have
\beq
\partial_{\mu}\sigma^{\mu}_{\alpha \dot \alpha} \psi^{* \dot
\alpha} =0.
\eeq
This has an immediate consequence:  if supersymmetry is to be
broken in some model, there had better be a light fermion which
can play the role of the Goldstino\cite{wittendsb}.

\noindent
{\bf Exercise:}  Prove the Goldstino theorem in generality.

\begin{figure}[htbp]
\centering
\centerline{\psfig{file=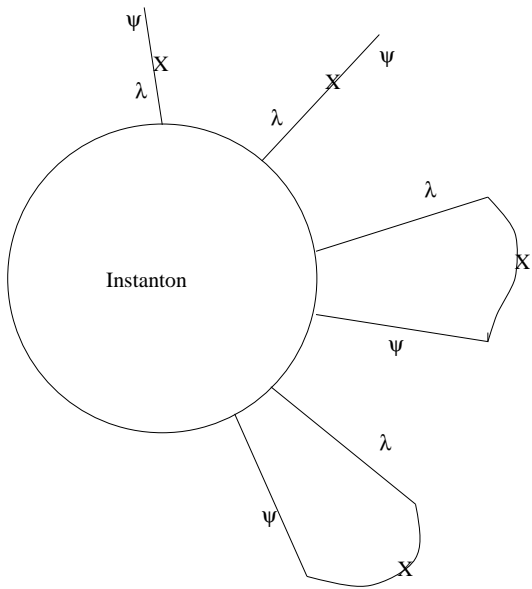,angle=-90,width=7cm}}
\caption{Instanton in supersymmetric QCD has four gaugino zero modes and two
quark zero modes at lowest order.  As indicated in the figure, the
scalar vev can be used to tie some of these together, generating
the two fermion term in the superpotential.}
\label{instantondiagram}
\end{figure}

Let us turn now to a particular example:  supersymmetric QCD with
gauge group $SU(N)$.  Consider first the case with no flavors,
$n_f=0$, i.e. a pure supersymmetric gauge theory.  The dynamical
fields are the gauge bosons and gauginos, $A^{\mu}$ and $\lambda$.
 The lagrangian is simply
\beq
{\cal L} = -{1 \over 4}F_{\mu \nu}^2 + i \lambda D_{\mu}\sigma^{\mu}
\lambda^*.
\eeq
Classically, this theory has no ground state degeneracy.
One expects
that the spectrum has a mass gap, like QCD, so there
is no candidate Goldstino and one does not expect
that supersymmetry is broken in this theory.
There is
a continuous global $U(1)$ (R) symmetry.  This symmetry, however,
is anomalous and only a discrete $Z_N$ subgroup survives in the
quantum theory.  This can
be seen, for example, by looking at instanton amplitudes.  An
instanton (fig. \ref{instantondiagram}) in this theory has $2N$ fermion zero modes,
meaning that there are expectation values for operators of the
form
\beq
\langle \lambda(x_1) \dots \lambda(x_{2N}) \rangle
\eeq
This breaks (explicitly) the $U(1)$ symmetry to a discrete
subgroup under which:
\beq
\lambda \rightarrow e^{2 \pi i \over N} \lambda.
\eeq
Just as quarks condense in QCD, it is reasonable to expect
that gluinos condense as well, i.e.
\beq
\langle \lambda \lambda \rangle = \Lambda^3.
\eeq
As we will see later, one can in fact {\it prove} that gaugino
condensation occurs in this theory.
Note, that $\lambda
\lambda$ is the lowest component of a chiral superfield,
$W_{\alpha} W^{\alpha}$, so the condensate is not an order
parameter for supersymmetry breaking, consistent with our
expectation that supersymmetry is unbroken.

Consider now the effect of adding ``quarks" to the theory, i.e.
fields $Q_f$ and $\bar Q_f$ in the $N$ and $\bar N$ representation of
$SU(N)$.  Here $f$ is a flavor index, $f=1, \dots,
N_f$. If these fields are massive, one does not expect
supersymmetry breaking, since the theory is much like the theory
with no quarks.  The massless case is more interesting.
\beq
V= {1 \over 2} \sum (D^a)^2
={\rm tr ~D^2}
\eeq
where $D_{ij}$ is the matrix:
\beq
D_{ij} = \hat D_{ij} - \delta_{ij} {\rm tr ~} \hat D~~~~~~
\hat D_{ij}=\sum_f Q_{if}^* Q_{jf} - \bar Q_{if} \bar Q_{jf}^*.
\eeq
Take, for
example, the case of two massless doublets in $SU(2)$, $Q$ and
$\bar Q$ (one ``flavor").  Classically there is a moduli space of vacua.
Consider the case $N_f<N_c$.  By a gauge
transformation, we can always take (writing $Q$ as a matrix in
flavor and color space)
\beq
Q = \left ( \matrix {v_1 & 0 & 0 & \dots & 0 \cr 0 & v_2 & 0 & \dots
& 0 \cr \dots \dots \dots \dots & 0 \cr 0 & 0 & 0 & \dots & v_{n_f}}
\right).
\eeq
Then the vanishing of $D_{ij}$ requires that
\beq
\bar Q = \left ( \matrix {v_1 e^{i \phi_1} & 0 & 0 & \dots & 0
 \cr 0 & v_2e^{i\phi_2} & 0 & \dots
& 0 \cr \dots \dots \dots \dots & 0 \cr 0 & 0 & 0 & \dots &
v_{n_f}e^{i\phi_{n_f}}} \right ).
\eeq
These expectation values correspond to moduli, which can be
described by the gauge-invariant fields
\beq
M_{f~ f^{\prime} }= \bar Q_f Q_{f^{\prime}}.
\eeq
The cases with $N_f \ge N_c$ are different.  For example, in the
case $N_f=N_c$, one has an additional solution, where $Q$ is
proportional to a unit matrix, and $\bar Q =0$.

\noindent
{\bf Exercise:}   Write the general flat direction for $N_f >
N_c$.   What gauge invariant fields are needed to describe it?

We would like to understand, quantum mechanically, what happens in
these flat directions.  Consider, first, the case of $SU(2)$ with
one flavor.  Here one has just the fields $Q$ and $\bar Q$; the flat
direction is the one we encountered in the MSSM, \beq Q = \left (
\matrix{0 \cr v} \right ) = e^{i\alpha} \bar Q. \eeq For non-zero
$v$, the $SU(2)$ gauge symmetry is completely broken. States
labeled by different $v$ are physically inequivalent. For example,
the masses of the gauge fields are different.  For large $v$, the
effective coupling is $g^2(v)$, and so the theory is weakly
coupled, and we should be able to analyze it completely. The
modulus which describes this flat direction we will call $\Phi$,
\beq \Phi = \bar Q Q. \eeq

To determine what happens to the flat
directions quantum mechanically, we should integrate out the
massive fields and study an effective action for $\Phi$.  Since
the theory is supersymmetric in the limit of weak coupling, we
expect this lagrangian to be supersymmetric.  This follows
from at least two considerations.  First, if one gauges the supersymmetry,
then one has a theory with a massless, spin-3/2 particle at low
energies.  Such a theory must be supersymmetric.  Alternatively,
one can make the argument purely in the global theory.
If supersymmetry is to be spontaneously broken, there must be a
massless fermion in the theory to play the role of the Goldstino.
Supersymmetry breaking corresponds, precisely, to the generation
of an $F$ or $D$-term for this field in the low energy lagrangian.

Given these statements, the low energy effective action has the
form,
\beq
{\cal L}_{eff} = \int d^4 \theta f(\Phi^{\dagger}\Phi)
+ \int d^2 \theta W(\Phi).
\eeq
Remarkably, the form of $W$ is completely determined by the
symmetries of the theory.  At a microscopic level, the theory has
a non-anomalous $U(1)_R$ symmetry, under which
\beq
Q \rightarrow e^{-i \alpha} Q ~~~~~~~
\bar Q \rightarrow e^{-i \alpha} \bar Q~~~~~~\theta \rightarrow e^{-i\alpha
\theta}.
\eeq
Under this symmetry, the $R$-charge of the quarks (i.e. the
fermionic components of $Q$ and $\bar Q$ is $-2$, and
the contribution to the anomaly cancels against that of the
gauginos with charge $1$.  $\Phi$ transforms
as
\beq
\Phi \rightarrow e^{-2i\alpha} \Phi
\eeq
The only holomorphic function, $W(\Phi)$,
with $R$-charge $2$ is:
\beq
W= {\Lambda^5 \over \Phi}.
\eeq
Here $\Lambda$ is the renormalization group scale of the
$SU(N)$ theory,
\beq
\Lambda= e^{-{8 \pi^2 \over b_o g^2}} = e^{-{8 \pi^2 \over 5g^2}}.
\eeq

As a check, we can determine the $\Lambda$ dependence from a
different argument.  We can describe the gauge coupling as the
background value of a chiral field, $S$,
\beq
-{1  \over 4} \int d^2 \theta S W_{\alpha}^2
\eeq
with
\beq
S = {1 \over g^2} + ia
\eeq
In the presence of the background field, the theory has an
additional symmetry:
\beq
S \rightarrow S + i {\alpha \over 16 \pi^2}
\label{stransformation}
\eeq
$$Q \rightarrow Q ~~~~~~~\bar Q \rightarrow \bar Q.$$
The rotation of the fermions cancels the shift of the lagrangian
from the shift in $S$.  $e^{-S}$ has $R$-charge $2$, so
\beq
W={e^{-S} \over \Phi}.
\eeq
This agrees with our expression above.

While the symmetry arguments are powerful, they may seem a bit
slick, and it is desirable to check that this term is in fact
present.  Since we are interested in the action of the
(at least approximately) massless states, it is appropriate
to study the Euclidean functional integral:
\beq
\int [d\phi] e^{-S}
\eeq
where $\phi$ refers to all of the various fields in the (full
microscopic) theory.  In the classical vacuum characterized by the
expectation value $v$, the field $\Phi$ is
\beq
\Phi = v^2 + v(\psi_Q + \psi_{\bar Q})\theta + \dots
\eeq
so the expected interaction includes terms such as
\beq
\Lambda^5 \int {d^2 \theta \over \Phi^2} = \dots
{\Lambda^5 \over v^4} \psi_Q \psi_{\bar Q}
\label{smallfluc}
\eeq
We wish to see if such terms appear in the path integral.

It is not hard to show that the interaction of eqn.
[\ref{smallfluc}] is generated by
instantons.  Instantons are Euclidean solutions of the classical
equations of motion\cite{coleman}.  They are expected to dominate
the Euclidean functional integral at weak coupling. Actually, a
simple scaling argument shows that there are no such solutions for
non-zero $v$; however, as 't Hooft explained in his original
paper on instantons\cite{thooftinstanton}, approximate solutions can be constructed.
Starting with the instanton of the pure gauge theory, which has
action $8 \pi^2 /g^2$ and a scale $\rho$, these solutions can be
constructed
perturbatively in $\rho v$.  Similarly, starting with the fermion
zero modes of the unperturbed solution (there are four associated
with the gauginos and two with the $Q$, $\bar Q$ fields), one can
construct two zero modes\cite{ads}.
The structure of the calculation is indicated in
fig. \ref{instantondiagram}.  In the figure, each of the lines
emerging from the blob denotes one of the unperturbed fermionic
zero modes; the scalar background is treated as a perturbation.
The actual calculation is straightforward,
and yields a non-zero coefficient for the expected operator\cite{pouliot}.
Other terms generated by the superpotential can be calculated as
well\cite{ads}.

The potential generated by the non-perturbative superpotential
in this model tends to zero at infinity.  In this regime, the
calculation is completely reliable.  To understand this, note
that at the microscopic level,
for large $v$, the theory consists of particles with mass
of order $v$, and the massless multiplet $\Phi$; there are no
direct couplings between the $\Phi$ fields themselves.
As a result, loop diagrams are dominated by physics
at the scale $M_V = gv$, and thus the effective
coupling is $g^2(v)$.  For sufficiently large $v$,
this coupling can be made arbitrarily small.
The Kahler potential receives only small corrections; the
superpotential, we have argued is exact (and in any case, the
semiclassical instanton calculation becomes more and more
reliable).  At strong coupling, it is conceivable that the Kahler
potential has some complicated behavior, leading, perhaps, to
a local minimum of the potential.  We do not have reliable methods
to explore this regime:  in this region of $\Phi$, it is not
even clear that $\Phi$ is the appropriate degree of freedom to
study.  On the other hand, for large $\Phi$, we have an
approximate moduli space.

Consider, now, the effect of adding a mass term, $m Q \bar Q$.
This breaks the $U(1)_R$ symmetry of the theory, leaving over
a $Z_2$.

\noindent
{\bf Exercise:}
Determine the symmetries of $SU(N)$ supersymmetric QCD with $N_f$ flavors
massless flavors.  Show that if all of the quarks have mass, the
theory has a $Z_N$ $R$-symmetry.
How do the $Q_{\alpha}$'s transform?

For small $m$, the superpotential is
\beq
W={ \Lambda^5 \over \Phi}  + m \Phi.
\label{wmass}
\eeq
The equation ${\partial W \over \partial \Phi}=0$ has two roots
\beq
\Phi = v^2 = \pm({\Lambda^5 \over m})^{1/2}.
\eeq
These two roots correspond to the spontaneous breaking of the
$Z_2$ symmetry of the theory.

As $m\rightarrow 0$, this computation is completely reliable, since
$v\rightarrow \infty$, so the coupling becomes weak.  It would
seem, that we could say nothing about stronger
coupling.  However, treating $m$ as a background field, we can
make a simple argument that the superpotential of eqn.
[\ref{wmass}] is exact.  We can assign charge $+4$ to $m$ under the
original $U(1)$ symmetry.  Then higher powers of $m$ must be
accompanied by higher powers of $\Phi$.  But perturbative and
non-perturbative contributions do not lead to such terms.
Alternatively, under the symmetry under which $\Phi$
is neutral and  $S$ transforms (eqn. [\ref{stransformation}]),
$m$ carries charge $2$. So higher powers of $m$ must come
with inverse powers of $\Lambda$.

So, for general $m$, we can compute the expectation value of the
superpotential:
\beq
\langle W \rangle = \pm m ({\Lambda^5 \over m})^{1/2}
\eeq
$$~~~~~~~= \pm(m\Lambda^5)^{1/2}.$$
But for large $m$, the theory is just pure gauge $SU(2)$.
The ``expectation value of the superpotential" in this theory
is naturally identified with
\beq
\int d^2 \theta W_{\alpha}^2 \approx \int d^2 \theta \langle
W_{\alpha}^2 \rangle = \int d^2 \theta \langle \lambda \lambda
\rangle = \Lambda_{LE}^3.
\eeq
In this expression,
$ \Lambda_{LE}$ is the $\Lambda$-parameter of the low energy
theory.  To determine the connection between this quantity and
the $\Lambda$ parameter of the full theory, note that
\beq
 \Lambda_{LE}= m e^{-{8 \pi^2 \over b_{LE} g^2(m)}}
 \eeq
 while
 \beq
 \Lambda= M e^{-{8\pi^2 \over b g^2(M)}}
 \eeq
 where $M$ is some high energy scale.
 $g^2(m)$ is determined from
 \beq
 8 \pi^2 g^{-2}(m) = 8 \pi^2 g^{-2}(M) + b \ln(m/M).
 \eeq
 Using $b_{LE}=6$, $b=5$ gives
\beq
\Lambda_{LE}^3 = (m \Lambda^5)^{1/2}.
\eeq
So the superpotential computed by these two different arguments
agree.  We can view this result in several ways.  First, the
calculation for small quark mass was completely reliable; we then
used holomorphy and symmetries to argue that the result was exact,
even for large quark mass.  The consistency of these computations
is a confirmation of these arguments.  Alternatively, we can
view the holomorphy arguments as reliable, and then argue
that we have {\it computed}
$\langle \lambda \lambda \rangle$ in QCD with a fermion in the
adjoint representation!

Models of this kind, with different gauge groups and various
numbers of flavors, exhibit a rich array of phenomena.  These are
reviewed, for example, in \cite{peskintasi}.
Of particular interest are the cases:
\begin{itemize}
\item $SU(N)$, $N_f <N$:  In these theories, as in our example
above, flat directions are lifted, and their is an approximate
moduli space at weak coupling.
\item $SU(N)$, $N_f =N$:  These theories have an exact moduli
space, but the quantum moduli space is not equal to the classical
one.
\item $SU(N)$, $N_f>N$:  In these theories, the quantum moduli
space is equivalent to the classical one.  These models exhibit
quite non-trivial dualities.
\end{itemize}

\subsection{Supersymmetry Breaking}

So far, we have seen that moduli can be lifted, but that
potentials in such cases tend to zero at weak coupling.  We might
hope to obtain ``real" supersymmetry breaking, i.e. supersymmetry
breaking with a unique, nicely behaved ground state.  The simplest
example of this phenomena is provided by the ``3-2
Model"\cite{peskintasi}.  This model has gauge group $SU(3) \times
SU(2)$, with fields $Q (3,2), \bar U (\bar 3,1), \bar D (\bar
3,1), L(1,2)$ (the numbers in parenthesis denote the $SU(3) \times
SU(2)$ representations), and superpotential
\beq
W= \lambda Q L \bar D.
\label{wthreetwo}
\eeq
With $\lambda = 0$, this theory has flat directions.  If the
$SU(3)$ coupling is large compared to the $SU(2)$ coupling, the
theory is like the $N_f = N-1$ theories, and instantons generate a
superpotential.  On the other hand, if $\lambda \ne 0$, there are
no flat directions.

\noindent
{\bf Exercise:}  Check that for non-zero $\lambda$, there are no
flat directions in the model.  Determine the two non-anomalous
global $U(1)$ symmetries of the theory. For the case $\Lambda_3
\gg \Lambda_2$, write down the superpotential which describes
the low energy theory and argue that it has a minimum with
broken supersymmetry.

\noindent
{\bf Exercise: (Advanced)}:  Consider the case $\Lambda_2 \gg
\Lambda_3$.  Show that with $\lambda=0$ this is a theory
with a quantum modified moduli space, and argue that with
$\lambda \ne 0$ supersymmetry is broken.

For small $\lambda$, the superpotential is the sum of the
non-perturbative superpotential and that of eqn. [\ref{wthreetwo}].
It is straightforward to show that this potential does not have a
supersymmetric minimum, and indeed has simply an isolated,
supersymmetry breaking minimum.  One can understand why this
happens by noting that any expectation values of the fields break
one or both of the non-anomalous $U(1)$ symmetries of the theory.
As a result, there are Goldstone bosons.  If supersymmetry is to
be unbroken, these Goldstone bosons must have superpartners. The
scalar superpartners would parameterize flat directions, but
there are no flat directions in the model.  This criterion, that
if there are broken symmetries in a theory without flat
directions, supersymmetry is broken, has proven useful for finding
examples of spontaneous supersymmetry breaking.
In recent years, many more examples of theories with supersymmetry
breaking have been exhibited, and other criteria for such
theories have been established.  This subject is thoroughly reviewed
in \cite{shirmanshadmi}.

\section{Supersymmetry Phenomenology and Model Building}

Before speculating about the dynamics of supersymmetry
breaking in nature, there
are some features of the MSSM
which we must discuss.  Not only are these features
important phenomenologically, but they are also
possible clues to the supersymmetry breaking
mechanism.  First, while the model has some formal
resemblance to the Standard Model, it has some important
differences.  One of the virtues of the SM is that there are no
renormalizable operators in the model which violate baryon number
or the separate lepton numbers.  As a result, it is not necessary
to impose these symmetries by hand, provided that the scales of
new physics are well above the scale of weak interactions.  This
is not the case in the MSSM.  There are dimension four operators,
such as
\beq
\int d^2 \theta[aQQQ+b\bar u \bar d \bar d+c QL\bar e],
\eeq
which violate these symmetries.  To suppress these, one
usually supposes that the model has an additional symmetry.  The
simplest hypothesis is that there is a discrete, $Z_2$, R
symmetry, known as $R$ parity, under which all of the ``ordinary"
particles (quarks, leptons, etc.) are invariant, while the ``new"
particles are odd.  This eliminates all of the dangerous couplings
above.  Weaker versions of this idea are possible, and lead to a
different phenomenology\cite{rparity}.  We will adopt the
$R$-parity violating hypothesis here, for simplicity.  It is
disappointing that from the start we must superpose some
additional symmetry to make this structure work, but we can take
comfort from the fact that such discrete symmetries are quite
common in string theory.  Even with this hypothesis, there are
other potential difficulties.  In the SM, the leading baryon
number violating operators are of dimension $6$.  However, even if
we suppress the dimension four operators in the MSSM, there are
potential problems from operators of dimension six, for example
\beq
QQQL,\bar u \bar u \bar d \bar e.
\eeq
With rather mild assumptions, these operators are highly
suppressed, however, and can be compatible with current
experimental bounds.

The R parity hypothesis has an important and desirable
consequence:  the lightest new particle predicted by supersymmetry
is stable.  This ``LSP" is, it turns out, a natural candidate for
the dark matter of the universe.  Cross sections for its
production and annihilation are such that one automatically
produces a density of order the closure density.

\begin{figure}[htbp]
\centering
\centerline{\psfig{file=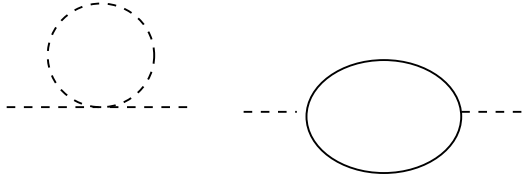,angle=-90,width=7cm}}
\caption{Infrared divergent contributions to the effective action.}
\label{massdiagrams}
\end{figure}

If we take the MSSM as our basic framework for thinking about
supersymmetry and nature, we obviously need to give masses to the
squarks, sleptons, gauginos and Higgs.  We could try to build some
supersymmetry-breaking model including these fields, along the
lines of the last section, but this turns out to be a challenging problem,
so we first adopt a simpler approach:  we just add masses for
these fields (along with certain cubic couplings of scalar
fields).  These ``soft breakings" don't reintroduce quadratic
divergences. In particular,
they don't introduce quadratic divergences.  This follows, more or
less, from dimensional analysis.  At very high energies,
corrections to masses should be proportional to the soft masses
themselves, and the resulting Feynman integrals should be
correspondingly less divergent.  Consider, for example, the
(massless) Wess-Zumino model, with $W= {\lambda \over 3 } \phi^3.$
The Yukawa and quartic couplings of this theory are:
\beq
{\cal L}_{int} = 2 \lambda \phi \psi \psi + {\rm c.c.} + \vert \lambda
\phi^2 \vert^2.
\eeq
At one loop there are two Feynman diagrams, indicated in fig. \ref{massdiagrams},
which contribute to the scalar mass:
\beq
\delta m^2 = \int^{\Lambda} {d^4 k \over (2\pi)^4}[{1\over k^2} - {1 \over
k^2}].
\eeq
Including a small supersymmetry-breaking mass for the scalars,
$m_s^2$, changes this expression to:
\beq
\delta m^2 = \int {d^4 k \over (2\pi)^4}[{1\over k^2+m_s^2} - {1 \over
k^2}]
\eeq
$$~~~~~~~=-{m_s^2 \ln(\Lambda^2 /m^2) \over 16 \pi^2}.$$
What is crucial, here, is that this is only logarithmically
sensitive to the cutoff $\Lambda$.

We can understand this result another way.  Introduce a field,
$X$, which we will refer to as a spurion.  This field is similar
in many respects to the background fields we have introduced in
earlier sections to describe coupling constants, and in practice
we may want to think of it as dynamical or we may not.  What is
crucial is that $F_X$ has a non-vanishing expectation value, i.e.
\beq
\langle X \rangle = \dots + \theta^2 \langle F_X \rangle.
\eeq
Consider, then, an operator of the form
\beq
{1 \over M^2} \int d^4 \theta X^{\dagger} X\phi^{\dagger} \phi.
\eeq
Integrating over $\theta$, this yields a mass term for the scalar
field,
\beq
{\langle F_X^{\dagger}
F_X \rangle \over M^2} \phi^{\dagger} \phi= m_s^2 \phi \phi.
\eeq
Note that this is a Kahler potential term, so it can (and will, in
general) be renormalized.  Working to quadratic order in $X$,
however, these renormalizations are at most logarihmic.  ($M$
should be thought of as comparable to the cutoff scale).

In terms of $X$, one can enumerate other possible soft breakings:
\begin{itemize}
\item
$
\int d^2 \theta {X \over M}m\phi \phi = m_s m \phi \phi + {\rm
c.c.}
$
\item
$
\int d^2 \theta  {X \over M} \phi \phi \phi + {\rm c.c.}
$
\item
$
\int d^2 \theta {X \over M} W_{\alpha}^2 = m_s \lambda \lambda +
{\rm c.c.}
$
\end{itemize}

To develop a phenomenology, then, we add to the MSSM soft
breaking terms corresponding to masses for squarks and sleptons,
as well as for gauginos.  We also add cubic couplings of Higgs
fields to squarks and sleptons.  We write the soft breaking lagrangian,
as:
\beq
{\cal L}_{soft}= \sum_{f,f^{\prime}}Q_f^*
(m_Q)^2_{f,f^{\prime}}Q_f^{\prime}
+
\sum_{f,f^{\prime}}\bar U_f^*
(m_U)^2_{f,f^{\prime}}\bar U_f^{\prime}
\label{soft}
\eeq
$$~~~~~~+ \sum_{f,f^{\prime}}\bar D_f^*
(m_D)^2_{f,f^{\prime}}\bar D_f^{\prime}
+ \sum_{f,f^{\prime}}\bar e_f^*
(m_E)^2_{f,f^{\prime}}\bar e_f^{\prime}$$
$$+ \sum_{f,f^{\prime}}L_f^*
(m_L)^2_{f,f^{\prime}} L_f^{\prime}$$
$$~~~~~~
+ \sum_{f,f^{\prime}} A^U_{f,f^{\prime}}Q_f \bar u_f^{\prime} H_U
+{\rm c.c.}
+ \sum_{f,f^{\prime}} A^D_{f,f^{\prime}}Q_f \bar d_f^{\prime} H_D
+{\rm c.c.}$$
$$~~~~~
+ \sum_{f,f^{\prime}} A^L_{f,f^{\prime}}L_f \bar e_f^{\prime} H_D
+{\rm c.c.}
+ \sum_i m_{\lambda_i} \lambda_i \lambda_i
$$
$$~~~~~ + m^2_{H_U} \vert H_U
\vert^2 + m^2_{H_D} \vert H_D
\vert^2 + B H_U H_D + {\rm c.c.} + \int d^2 \theta \mu H_U H_D.$$

If nature {\it is} supersymmetric, determining these parameters
and understanding their origin will be a principle goal of
particle physics.  The first question one might ask is:  how many
parameters are there?  It is easiest to count relative to the
Standard Model.    There, in defining the usual KM phases, one has
already used most of the freedom to make field redefinitions.  In
order to count the remaining parameters, one must ask what is the
remaining freedom.  Before considering ${\cal L}_{soft}$, the
theory has several global $U(1)$ symmetries. These are
\begin{itemize}
\item
An R symmetry which rotates all the fields, under which one can
define the $R$ charge of the Higgs fields to be two, and of all
other matter fields to be zero.
\item
Peccei-Quinn symmetry (not an R symmetry):
\beq
H_U\rightarrow e^{i \alpha} H_U~~~~~~~~H_D \rightarrow e^{i\alpha}H_D
\eeq
$$Q \bar u\rightarrow e^{-i\alpha} Q \bar u~~~~~Q \bar d \rightarrow
e^{-i \alpha} Q \bar d$$
\item
Three lepton number symmetries.
\end{itemize}
So all together there are five global
symmetries.  One of these (the
overall lepton number) is preserved by ${\cal L}_{soft}$, so we
have the freedom to make four redefinitions.  Now we can count.
Each of the five matrices, $m_Q^2, m_{\bar U}^2$, etc., is a $3\times
3$ Hermitian matrix, with nine parameters.  $A_U$, $A_D$ and $A_L$
are general complex matrices with $18$ parameters.  The three
gaugino masses (complex) represent six additional parameters.  In
the Higgs sector, we have six more real parameters.  So all
together, there are $111$ parameters, from which we must subtract
four possible redefinitions, and the two Higgs parameters of the
usual Standard Model.  {\it This leaves 105 new parameters.}

The parameter space of the MSSM is enormous.  It is possible
that at some point we will have a compelling theory which will predict the
values of these quantities, but this is not the case today.
To explore this
space, we need to make some hypotheses.
Perhaps the simplest possibility is to assume
some simple structure for the soft-breaking masses at some very
high energy scale.
The following relations are often referred to as the ``MSSM"
or the ``supergravity model:"
\beq
m_Q^2 =m_{\bar U}^2 = m_{\bar D}^2 = m_o^2
\label{universality}
\eeq
$$m_{\lambda_i} = m_{1/2}$$
$$A_U=A_D=A_L = A,$$
i.e. all of the various matrices are supposed proportional to the
unit matrix.  These choices have several virtues.  First, one can
do phenomenology with a small set of parameters.  Second, the
various flavor-changing neutral current constraints, which we will
discuss below, are automatically satisfied.  Experiments,
especially at LEP and the Tevatron rule out much of this parameter
space.  TeVII and the LHC will explore much of what remains.

One of the questions we will address in the remainder of this
lecture is:  ``How plausible is this structure."
Before doing so, however, it is important to mention the other
prediction of this framework, apart from dark matter:  Coupling
Unification.  If we assume, consistent with the hierarchy problem,
that the susy thresholds are at scales of order $100$'s of GeV,
and if we run the observed gauge couplings from their values at
$M_Z$, one finds that the couplings unify.  This
unification occurs at a scale of order $2 \times
10^{16} ~{\rm GeV}$.  It can be thought of as a prediction of
$\alpha_s$ given $\alpha_W$ and $\alpha$, and this prediction is
good to about $2\%$.

\noindent
{\bf Exercise:}
These results are quite easy to derive.
At one loop, if the couplings, $\alpha_i$, are equal at a scale
$M_{GUT}$, one has
\beq
2 \pi \alpha_{i}^{-1}(M_Z) = 2 \pi \alpha_{i}^{-1}(M_{GUT})
+ b_o^i \ln(M_Z/M_{GUT}).
\eeq
If unification is in $SU(5)$, then the hypercharge of the standard
model, $Y$, is proportional to an $SU(5)$ generator, $\tilde Y$,
which is normalized just like any $SU(N)$ generator:
\beq
{\rm tr ~}(\tilde Y^2) = {1 \over 2}.
\eeq
Compare this with the conventional values of the hypercharge,
to show that
at $M_{GUT}$, the gauge couplings are related by
\beq
\sin^2(\theta_W) = {3 \over 5}.
\eeq
Using the renormalization group equations, show that the
couplings are equal at a scale of about $2 \times 10^{16}$ GeV.

\begin{figure}[htbp]
\centering
\centerline{\psfig{file=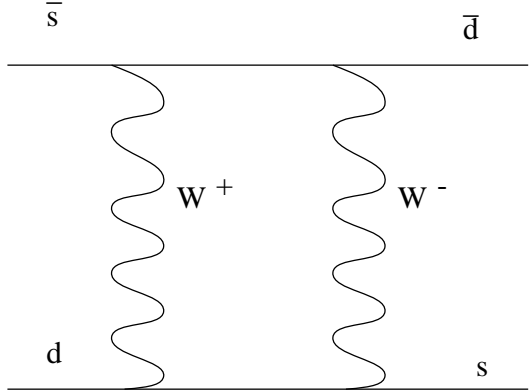,angle=-90,width=7cm}}
\caption{A contribution in the standard model to $K -\bar K$ mixing.}
\label{standardbox}
\end{figure}

The simple hypothesis of eqn. [\ref{universality}] for the soft breaking masses
avoids a number of potential problems.  Apart from proton decay,
it also means that separate lepton numbers are conserved;
off-diagonal elements in the matrices
$A^L,m_E^2,m_L^2$ in eqn. [\ref{soft}] (in the basis where the
charged lepton masses are diagonal) could lead to $\mu \rightarrow
e \gamma$, for example.  Other rare processes are also suppressed.
$K -\bar K$ mixing is a tiny effect, which the SM
predicts at roughly the correct level (fig. \ref{standardbox}).  In the SM,
\beq
{\cal L}_{eff} \propto {\alpha_W \over 4 \pi} {m_c^2 \over
M_W^2} G_f \sin^2(\theta_c) \ln(m_c^2 / m_u^2){\cal O}.
\eeq
${\cal O}$ is a certain four-fermi operator which violates
strangeness by two units.
This formula is meant to show the various sources of suppression.
The GIM mechanism leads to a suppression by a factor of
$m_c^2/M_W^2$; there is further suppression by Cabbibo angles.
The matrix elements of ${\cal O}$ are also suppressed by
powers of $m_K^2$.

\begin{figure}[htbp]
\centering
\centerline{\psfig{file=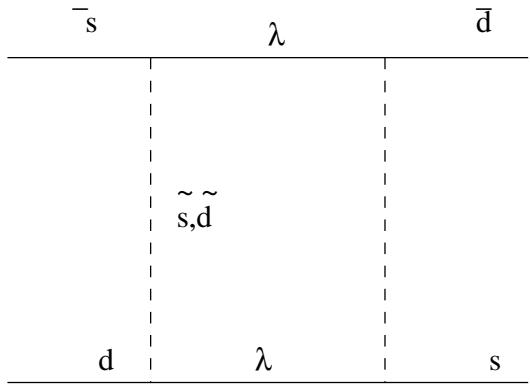,angle=-90,width=7cm}}
\caption{A potentially large contribution to $K -\bar K$ mixing
in supersymmetric models.}
\label{susybox}
\end{figure}

In supersymmetry, there are new contributions,
which are potentially quite large.
An example is indicated in fig. \ref{susybox}.  In general, this graph
has no suppression by $m_c^2 \over M_W^2$; it involves $\alpha_s$
rather than $\alpha_W$ (a factor of about $10$ in the rate); the
chiral symmetry suppression is absent (i.e. the factor
of $m_K^2$).  If there are no special
cancellations, one requires that the scale of supersymmetry
breaking be of order $100$'s of TeV to adequately suppress this
graph.  But cancellations are automatic if the hypothesis of
eqn. [\ref{universality}] is satisfied.  Some degree of degeneracy seems
almost inevitable if one is to understand these facts.

\begin{figure}[htbp]
\centering
\centerline{\psfig{file=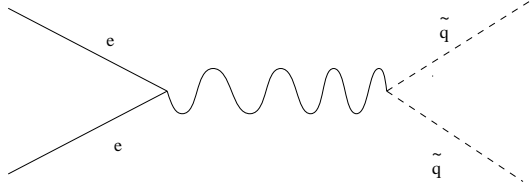,angle=-90,width=7cm}}
\caption{Diagram involving $\gamma$ or $Z$ exchange giving rise
to squark and slepton production in $e^+ e^-$ annihilation.}
\label{squarkproduction}
\end{figure}

Apart from the constraints on soft masses which come from these
rare processes, there are by now significant constraints from
direct searches.  I won't review these in detail here, but one
has, roughly:
\begin{itemize}
\item
Gluinos:  $m_{\lambda} > 225 ~{\rm GeV}$
\item
Neutralinos (the lightest neutral fermions from the
Higgsino/wino/zino sector)
$m_{\chi^o} > 30 ~{\rm GeV}$
\item
Charginos:  $m_{\chi^{\pm}}> 100-150 ~{\rm GeV}$
\item
Sleptons:  $m_{\tilde e} > 90 {\rm}~ GeV$
\end{itemize}
An example of a process leading to selectron production in
$e^+ e^-$ annihilation is indicated in fig. \ref{squarkproduction}.

\section{Approaches to Understanding Soft Breakings}

While it is good to have a parameterization of supersymmetry
breaking, one might hope to understand these phenomena at some
more fundamental level, and to make predictions for these soft
breaking parameters.  There are a few ideas about how
supersymmetry might be broken in nature, and about how this
breaking is manifest in the spectrum of the MSSM or some
generalization.  In this section, I will briefly review these.
As we will see, while any of these ideas might be
correct, none, as they currently stand are
completely compelling.  Each has a possible realization in string
theory.

\subsection{``Supergravity" Breaking}

So far, we have treated supersymmetry as a global symmetry.  In a
theory with gravity (and in particular in string theory), we
expect that it should be a local symmetry.  In other words, under
supersymmetry, the graviton should have a fermionic partner, of
spin $3/2$, the gravitino, $\psi_{\mu \alpha}$.  In $N=1$
supergravity, terms with up to two derivatives in
the effective lagrangian are specified, as in
global supersymmetry, by three functions, the Kahler potential,
$K(\phi_i,\phi_i^{\dagger})$, and two (sets of) holomorphic
functions, the superpotential, $W$, and the gauge coupling
functions, $f_a$.  The scalar potential is given by
\beq
V= e^K[({\partial W \over \partial \phi_i} + {\partial K \over
\partial \phi_i} W) g^{i \bar j}
({\partial W^* \over \partial \phi_j^*} + {\partial K \over
\partial \phi_j^*} W^*) - 3 \vert W \vert^2].
\label{supergravity}
\eeq
In this expression, the Kahler metric, $g_{i \bar j}$ is given by
\beq
g_{i \bar j} = \partial_i \partial_{\bar j} K.
\eeq

The reasons for putting ``supergravity" in quotes in the title of
this section are twofold.  First, because supergravity theories
are not renormalizable theories, they are at best low energy
descriptions of some more fundamental theory (string theory).
Second, what are usually called supergravity models
actually involve a quite specific set of assumptions about the
structure of the Kahler potential (and often the functions $f_a$).
In particular, these models assume that the various chiral fields
fall into two types, called ``visible sector" and ``hidden sector"
fields, and which we will denote by $y_i$ and $z_A$, respectively.
The $y_i$'s
include the quarks, leptons and Higgs, while the $z_A$'s are
called ``hidden sector" fields, and are supposed to be associated
with supersymmetry breaking.  The $z_A$'s are assumed to
be responsible for supersymmetry breaking.  The superpotential is supposed to
break up into two pieces:
\beq
W=W(y_i) + W(z_A).
\eeq
This is a plausible assumption, which, given the holormorphy of
$W$, might be enforced by symmetries, at least at the level of
relatively low dimension operators.  The Kahler potential is also
supposed to break up in such a fashion:
\beq
K = \sum_i y_i^{\dagger} y_i + \sum_A z_A^{\dagger} z_A.
\label{canonicalkahler}
\eeq
This latter assumption yields universality.  It is often said to
follow from the universality of gravity, but this is clearly not
the case; no symmetry forbids a more complicated
form.  Generically, this
doesn't hold in string theory (though it is sometimes true in some
approximation).

\noindent
{\bf Exercise:}
An example of this sort of model is provided by the ``Polonyi"
model.  Here one just has a single hidden sector field, $z$.  The
superpotential is given by
\beq
W_{hid} = m^2(z+ \beta).
\eeq
Show that for
$\beta = (2 + \sqrt{3}M)$, where $M$ is the reduced Planck
mass,
$$G_N= {1 \over 8 \pi M^2},$$
the potential has a minimum for
$V=0$, with
$$Z= (\sqrt{3}-1)M ~~~~~~m_o^2 = 2 \sqrt{3} m_{3/2}^2~~~~~
A=(3-\sqrt{3})m_{3/2}~~~~~m_{3/2} = {m^2 \over M}
e^{(\sqrt{3}-1)^2/2}$$
Here $m_{3/2}$ is the gravitino mass.  Gaugino masses
in such models are assumed to arise from a coupling
\beq
\int d^2 \theta Z W_{\alpha}^2.
\eeq

\subsection{Incorporating Dynamical Supersymmetry Breaking}

One would like to understand supersymmetry breaking dynamically.
Within the framework of gravity mediation, one can suppose that
the hidden sector fields, $z_A$, correspond to some
supersymmetry-breaking theory, such as the $3-2$ model discussed
earlier.  Assuming that the Kahler potential has, say, the form of
eqn. [\ref{canonicalkahler}],
one can work out the scalar spectrum in detail, and it
is not qualitatively different than that of the Polonyi model
discussed above.  There is, however, a problem when one discusses
the masses of the gauginos.  In models in which the hidden sector
is dynamical, operators of the type
\beq
\int d^2 \theta f(\phi) W_{\alpha}^2
\eeq
are typically highly suppressed.  In the $3-2$ model, for example,
the operator $f$ of lowest dimension is
$QL\bar d \over M_p^3$.  This leads to a gaugino mass of order
\beq
m_{1/2} \sim {\Lambda^4 \over M_p^3}
\eeq
as opposed to scalar masses of order
\beq
m_o^2 \sim {\Lambda^4 \over M_p^2}.
\eeq
Generically, these are far too small.

Recently, a solution has been proposed to this problem, referred
to as ``anomaly mediation"\cite{anomalymediation}.
Suppose that the Kahler potential is not that of eqn [\ref{canonicalkahler}],
but instead has the property that the scalar masses vanish to some
degree of approximation.  Then there are contributions to both
scalar and gaugino masses proportional to gauge couplings.  These
are associated with certain anomalous transformations in the
theory, which require a modification of the argument above that
fermion masses are extremely small.  One finds (here $\alpha$ is
the unified coupling at the high energy scale)
\beq
m_{1/2} \sim ({\alpha \over 4 \pi})m_o
\eeq
while the scalar masses are given by
\beq
m_{\tilde q}^2 = {1 \over 32 \pi^2} c_o b_o g^4 m_o^2
\eeq
where $b_o$ is the lowest order contribution to the $\beta$
function and $c_o$ is related to the anomalous dimension of the
scalar field through
\beq
\gamma = c_o {g^2 \over 16 \pi^2}.
\eeq

Examining these formulas, it is clear that there are two issues
which must be addressed.  First, one needs to understand why the
scalar masses are so much smaller than the value one naively
expects.  Second, the simplest formula predicts that some of the
scalar masses are negative, leading to breaking of electric
charge.  A number of ideas have been proposed for resolving both
questions\cite{recentanomalymediation}.

\begin{figure}[htbp]
\centering
\centerline{\psfig{file=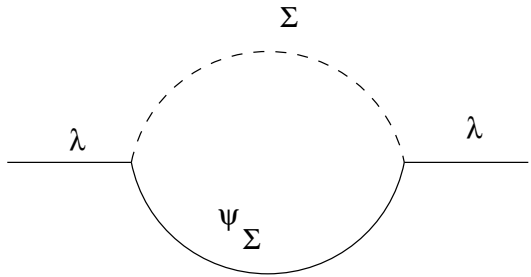,angle=-90,width=7cm}}
\caption{One loop diagrams which give rise to ``anomaly mediation."  $\Sigma$
denotes a heavy chiral field.}
\label{anomalymed}
\end{figure}

To understand what goes wrong with the naive arguments for the
gaugino masses requires careful consideration of various issues in
supergravity theories, but it can be understood in the following
heuristic way.  Suppose that in the theory there is a very
massive chiral multiplet.  The lagrangian for this field includes
the term
\beq
\int d^2 \theta M \Phi^2 ,
\eeq
as well as a soft-breaking ``B-term,"
\beq
m_{3/2}M \Phi^2 + {\rm c.c.}
\eeq

So examine the diagram of \ref{anomalymed}, where a gaugino couples to
the heavy field.  The diagram gives a result proportional to
${\alpha \over \pi} m_{3/2}$.
But this already violates our earlier arguments about
gaugino masses.  To understand what went
wrong, note first that this diagram is independent
of the mass of the heavy field.  Suppose now one introduces a
Pauli-Villars regulator with mass $\Lambda$.  Because the diagram
with the Pauli Villars field, like the original
diagram, is mass independent, it survives in the
limit $\Lambda \rightarrow \infty$ and yields overall
a vanishing gaugino mass.  This is in accord with the symmetry argument.  On the
other hand, for a {\it massless} field, the Pauli-Villars term now
introduces a contribution, which agrees with the anomaly mediation
formula.  When this observation was first made in
\cite{macintire}, its theoretical underpinnings
were not pursued, as it did not seem of great importance; the mass
seemed too small.  Fermion masses would be loop suppressed
relative to scalar masses.  As we have noted, the authors of
\cite{anomalymediation,recentanomalymediation}, in addition
to putting the theory on a clearer footing, have suggested
some scenarios where a suitable spectrum might naturally
result.

\subsection{Gauge Mediation}

\begin{figure}[htbp]
\centering
\centerline{\psfig{file=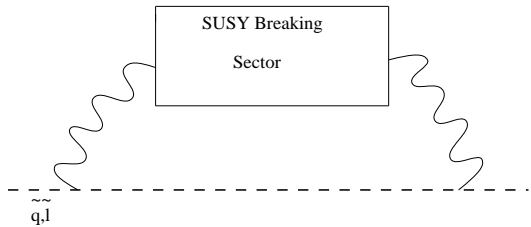,angle=-90,width=7cm}}
\caption{The general structure of gauge-mediated models.
The box indicates dynamics at the supersymmetry breaking scale.  The
wavy lines indicate various types of gauge exchanges, including
gauginos and scalars.}
\label{mgmschematic}
\end{figure}

As an alternative to the gravity mediation hypothesis, which has
been discussed in the previous section, suppose that supersymmetry
is broken at lower energies, smaller than $\sqrt{M_ZM_P}$.  For
reasons that will become clear shortly, we will refer to this
possibility as ``gauge mediation"\cite{gaugemediation}.  The basic model building
strategy is indicated in fig. \ref{mgmschematic}.  We will not review these
models here, but simply mention the basic ideas, and refer
the interested reader to some of the excellent reviews which
are now available\cite{gmreviews}.  One supposes that
supersymmetry is broken by some set of fields.  Some of these
fields carry ordinary gauge quantum numbers, so that masses of
squarks, sleptons, and gauginos can arise through loops involving
ordinary gauge fields (and their superpartners).
A typical mass formula (associated with the case of ``minimal
gauge mediation") has the form:
\beq
\tilde m^2 = 2 \Lambda^2[c_3 ({\alpha_3 \over 4 \pi})^2+
c_2 ({\alpha_2 \over 4 \pi})^2+
{5 \over 3} ({Y \over 2})^2 ({\alpha_1 \over 4 \pi})^2]
\label{gmspectrum}
\eeq
$$~~~~~m_{\lambda_i}=c_i {\alpha_i \over 4 \pi} \Lambda.$$
Here the parameter $\Lambda$ is typically the ratio of the
Goldstino decay constant ($F$, with dimensions of mass-squared),
to the energy scale of the susy-breaking interactions, $M_{sb}$,
$\Lambda = F/M_{sb}$.
Such an approach
automatically satisfies the constraints imposed by strangeness
changing neutral currents, since the masses are, to a good
approximation, only functions of gauge quantum numbers.

Because their dynamics is controlled by renormalizable
interactions at low energies, gauge-mediated models tend
to be highly predictive.  Such models typically have only
a small number of parameters beyond those of the SM.  The
phenomenology of these theories is also quite distinctive.
In particular, the gravitino is far lighter than expected
in gravity-mediated models.  Indeed, examining eqn.
[\ref{gmspectrum}], one sees that the parameter $\Lambda$, and thus
the scale of these new interactions and of the Goldstino decay
constant, could be as low as $10 ~{\rm TeV}$.  In this case, the
gravitino mass is of order
\beq
m_{3/2} \sim {(10 ~{\rm TeV})^2 \over M_p}
\eeq
$$~~~~~~~\sim 0.1 ~{\rm eV}.$$
The couplings of this particle are suppressed only by the $10
~{\rm TeV}$ scale, rather than the Planck scale.  Moreover, this
particle is now the LSP.  The next to lightest supersymmetric
particle now may be charged or neutral, in principle.  It can
decay with a track length as short as a fraction of a cm to final
states containing an (unobserved) gravitino\cite{ddrt}.

The detailed phenomenology of these models is quite rich, and
model building possibilities have only been partially explored.  It should
be noted that some of the most elegant models which have been
constructed so far have supersymmetry broken at a scale much
larger than $10~ {\rm TeV}$\cite{yael}.  Others have features such as
composite quarks and/or Higgs fields.
No model is yet totally compelling by itself, but it seems
possible that a truly ``Standard" supersymmetric model might
emerge from this framework.

\section{String/M Theory Phenomenology}

If we are trying to develop a superstring phenomenology, the first
question we might ask is:  what is string theory?  As you know, we
have good reason to believe that there is some overarching theory,
various limits of which look like weakly coupled string theory, or
eleven dimensional supergravity, or other things.  All we really
understand, however, is what the theory looks like on certain
moduli spaces with a great deal of supersymmetry.  It is highly
unlikely that the state of this theory which describes nature sits
on this moduli space.  All thinking about string phenomenology to
date assumes that this state is a stationary point of some
potential in an {\it approximate} moduli space. We will see that we have
some understanding of these approximate moduli spaces as well.
However, even this need not be correct.
It could be that the ground state is some truly isolated
point.  In that case, the question:  what is string theory?
assumes greater urgency.  It is not clear that this state
need be connected, in any way, to the states we understand, nor that
there is any small parameter which might allow us to
study this state.  Saying simply that ``string theory is the theory of quantum
gravity," in this situation, will have little
content\footnote{I thank Leonard Susskind for conversations which
sharpened this question.}.

We won't offer any general answers to this question here.  Instead,
we will assume that nature is approximately supersymmetric, and
that indeed we sit at a point in some approximate moduli space.
We now understand much about supersymmetry dynamics and
phenomenology.  In the remaining lectures we will apply this
understanding to String/M theory.  We will use supersymmetry both
to constrain possible dynamics, and to consider phenomenology.

Apart from hierarchy, we
will see some reasons to believe that low energy
supersymmetry might play some role in string
theory, but these will not be compelling.  The question of whether
string theory predicts low energy supersymmetry, or something
dramatically different, is the most
important question of string phenomenology, and the reader should
keep this in mind throughout.
There was, for a long time, a prejudice that string theory and low energy
supersymmetry somehow go together.  Recently, there have been
proposals to solve the hierarchy problem with large extra
dimensions\cite{largeradii,precursors} or warped
extra dimensions\cite{rs}.
In these proposals,
the standard model lives on a brane or wall of some sort,
and gravity propagates in the bulk.  The first set of proposals
generally invoke bulk supersymmetry as part of the explanation
of the hierarchy, but typically there is no relic of
supersymmetry at low energies in the conventional sense.  The second set of proposals
don't seem to require supersymmetry at all\cite{goldbergerwise}!
These lectures may
reinforce the prejudice that supersymmetry
is important, mainly
because it will be easiest to analyze states which
are supersymmetric, in some approximation.  Indeed, it
has been hard, up to now, to relate
the extra dimension proposals to detailed
constructions in string theory (with the exception
of \cite{ibanezlarge}).  It is unclear
what the parameters of these constructions are, and whether
any of these would permit a systematic computation
of physical quantities, i.e. would permit
the prediction of physical quantities directly
from some underlying (string) theory. In the framework
of low energy supersymmetry will see that, even though
the true vacuum of $M$-theory cannot be described in a systematic
weak coupling expansion, there may be a small parameter, which would
allow computation of some quantities.  Perhaps, at the moment, the
best thing we can say is that this picture is consistent with
things we know about nature, and might permit some predictions.
It is your challenge to do better.

\subsection{Weakly Coupled Strings}

We
would like first to get some feeling for the moduli spaces of
string/M theory.  We begin by considering the heterotic string
theory, compactified on $T^6$.  This theory has $16$
supersymmetries ($N=4$ in four dimensional counting).  It is not hard
to see how the ten
dimensional fields fall into $N=4$ multiplets.  Vector
indices decompose as four dimensional Minkowski indices and
internal indices, associated with the $6$ dimensional internal
space.  We have seen that the two-derivative terms of $N=4$
theories respect an $SU(4)$ symmetry; this is just inherited from
the $O(6)$ symmetry of the compact dimensions, at infinite radius.
Clearly it is not exact; it is broken by the finite size of the
torus, and these effects will show up at the level of higher
derivative operators.  This $SU(4)$ {\it is} convenient for
classifying fields.  The vector fields of $10$ dimensions, for
example, decompose as four dimensional vectors and scalars:
\beq
A_M \rightarrow A_{\mu},\phi_i, i=1 \dots 6.
\eeq
The gauginos decompose as
\beq
\lambda^A \rightarrow \lambda^{\alpha,i} ~~~i=1,\dots 4
\eeq
where $\alpha$ is a space-time spinor index and $i$ an $SU(4)$
index.  The supersymmetry generators have a similar decomposition.
 The metric decomposes into the four dimensional metric, $6$
 vectors, and $21$ scalars; the antisymmetric tensor of ten
 dimensions decomposes as a four dimensional antisymmetric tensor
 (dual to a scalar), $6$ vectors, and $15$ scalars.  The
 gravitino decomposes into four four-dimensional gravitinos,
 $\psi_{\mu}^{\alpha,i}$, and $24$ spinors, $\psi_{j}^{\alpha,i}$.
 There is also a scalar from the ten-dimensional dilaton.
To see how these states fit into $N=4$ multiplets, note first that
the ten-dimensional vectors give four dimensional vector
multiplets.  The four dimensional supergravity multiplet contains
a graviton, four gravitinos, $6$ vectors, four Weyl spinors, and
two scalars (see \cite{polchinski} for example). So the state
counting is correct to correspond to a supergravity multiplet and
$6$ additional vector multiplets.

What does characterize the theory we call string theory or
$M$-theory is the existence of limits in which there is a
perturbation expansion corresponding to a weakly coupled string
theory.  In each case, one can identify the modulus which
corresponds to the string expansion parameter in a variety of
ways.  For the heterotic case, one can examine the low energy
effective lagrangian.  This lagrangian can be presented
in a variety of ways.  One has the freedom to redefine the metric
by a Weyl-rescaling, i.e. by $g_{MN} \rightarrow f(\phi) g_{MN}$,
where $\phi$ is the dilaton field.  A particularly convenient
choice is the ``string metric," in which the
ten-dimensional action has the form:
 \beq
\int d^{10} x \sqrt{g} e^{2 \phi}[-{1 \over 4} F_{MN}^2 + i
\lambda D_M \Gamma^M \lambda + {\cal R} + \dots] \eeq
Here the dilaton appears out in front of the
action (``string frame"), and the unit of length is $\ell_s$, the
string scale.  This is also the cutoff scale for this effective
lagrangian.  As a result, $e^{-2\phi}$ plays the
role of a dimensionless expansion parameter.  $\phi \rightarrow
\infty$ corresponds to weak coupling; $\phi \rightarrow +\infty$
to strong coupling.  That $\phi$ plays the role of the string
coupling can also be seen directly in string theory, by studying
the conditions for conformal invariance, for example.  A more
conventional presentation is provided by the Einstein metric (Einstein
Frame), in
which the curvature term in the action is independent of the
dilaton.

{\bf Exercise:}  Determine the transformation between the string
metric and the Einstein metric.

If we now compactify this theory on a six torus, with
\beq
\int d^6 y \sqrt{g} =V
\eeq
then the four dimensional coupling is
\beq
g_4^{-2} = e^{2\phi} V.
\eeq

These compactifications have many interesting features.  For
example, they exhibit electric-magnetic duality, as well as a
duality to type II theory compactified on $K_3 \times T_2$.
However, for our discussion, one feature is particularly striking:
this moduli space is exact,both perturbatively and
non-perturbatively.  This follows, as in the case of field theory,
by studying the constraints imposed by supersymmetry on the low
energy effective action.  This is a generic feature of $N \ge 2$
supersymmetry in four dimensions, and $N \ge 1$ in higher
dimension.  Thus we have our first observation relevant to
phenomenology:  There may be some states of string theory which
resemble our world, but there are certainly many which do not!

One of the interesting features of the moduli space is the
existence of various duality symmetries which relate different
points.  One of the most familiar is $T$-duality, under which, in
the case of compactification on a circle of radius $R$, $R
\rightarrow {1 \over R}$.  In the case of the heterotic
string, at the self-dual point, the gauge
symmetry is enlarged to $SU(2)$.   At the enhanced symmetry point,
$R$ is a component of an $SU(2)$ triplet (or
more properly, $\delta R$ where $R= \sqrt{2} + \delta R$).
Since $\delta R$ changes sign under the symmetry, this can be
identified with a gauge transformation, a rotation by $\pi$ about
the $y$ axis in isospace.  Indeed, all of the $T$-dualities of the
weak coupling limit can be identified as gauge transformations.

Electric-magnetic duality, or
$S$-duality, exchanges $g^2 \rightarrow 1/g^2$.
This
symmetry can be understood in a variety of ways.  For example,
under the duality which connects the heterotic string theory on
$T^4 \times T^2$ and Type II theory on $K_3 \times T_2$, $S$
duality is mapped to $T$ duality\cite{wittenusc}. For a particular
choice of $g^2$, one has an unbroken discrete symmetry which
exchanges $\vec E$ and $\vec B$.

It is interesting that one can find points of
enhanced symmetry under which all of the moduli transform.
The simplest example of such a {\it maximally enhanced
symmetry} is provided by the IIB theory in ten dimensions.  There,
one has an $SL(2,Z)$ symmetry, one of whose generators
transforms the dilaton as
multiplet, $\tau = a + {i \over g^2}$, as
\beq
\tau \rightarrow -{1 \over \tau}.
\eeq
For a special value of $g$, this symmetry is restored, and since
this is the only modulus, all of the moduli transform.  Such
enhanced symmetry points, in theories with less supersymmetry, are
potentially interesting for a variety
of reasons.  First, they are automatically stationary points of
whatever may be the effective action, so they are candidate ground
states.  They are also of interesting, as we will discuss later,
for various issues in
cosmology.

What about four dimensional compactifications with $N=1$
supersymmetry?  Based on our field theory experience, we might
expect that the classical moduli are not moduli; at best,
there are approximate moduli for weak couplings and/or large
compactification radii.  We consider this question first in the
context of Calabi-Yau compactifications of the
heterotic string at large radius(see  \cite{gsw,greenetasi}).
For large $R$, one constructs solutions perturbatively in
$\alpha^{\prime}R^2$, either by looking for renormalization
group fixed points in two dimensions, or by solving the field
equations in the effective field theory.  In the two dimensional
description, one has equations for conformal invariance, constructed
order by order in $\alpha^{\prime}$.  In the ten dimensional
description, one first solves the equations including operators
with at most two derivatives, and then considers the effect
of higher derivative operators.  One finds that these schemes
are equivalent, and that to lowest order,
Ricci-flat backgrounds along with
appropriate gauge field backgrounds solve the equations of motion.

The first question one might ask is:  under what circumstances can
these lowest order solutions be generalized to exact solutions?
Are the corresponding moduli found at lowest order truly moduli of
the classical theory?  And what is their quantum mechanical fate?
These questions can both be addressed by considering the low
energy, {\it four dimensional} field theory.
Consider, first, the question of
constructing solutions given a lowest order solution.  What
is the issue?  Suppose one has solved the classical equations of
one of the superstring theories to lowest order in
$\alpha^{\prime}$, or more precisely $\alpha^{\prime}/R^2$, and
suppose that supersymmetry is preserved to this order.  Then the
spectrum of fluctuations about this background includes states
with mass of order $1/R$, mass of order $\ell_s$ (the string
scale), and some finite number of states of zero mass.  The
question of finding a solution is the question of whether there are
tadpoles for these states in higher orders of approximation.

For the massive states, tadpoles do not represent obstructions to
finding solutions.   This is already clear in simple field
theories.  If
\beq
{\cal L}= m^2 \phi^2 + \Gamma \phi,
\eeq
the massive field simply shifts so as to cancel the
tadpole.  In the two dimensional, conformal field
theory description, this is
just the statement that if one adds a small term proportional to
an irrelevant operator, the system flows to the fixed point.
However, for massless states, tadpoles are potentially more
serious.  There is no guarantee that one can find a
(static) solution, and in general one cannot.
But this statement also makes clear that to investigate the
existence of solutions, one should integrate out the massive
fields and examine the low energy effective lagrangian.  This
effective lagrangian must be supersymmetric and respect the
various symmetries, and as usual, this provides powerful
constraints.

In constructing weak-coupling compactifications of the heterotic
string, one must not only choose, at lowest order, a Ricci-flat
background for the gravitational field, but one must also choose a
background gauge field.  The general problem is discussed in
\cite{gsw,greenetasi}.  The simplest choice is to take the gauge
field to be in an $SU(3)$ subgroup of the gauge group, and to set
this field equal to the spin connection.
In the $\alpha^{\prime}$ expansion for these
configurations, there is then a simple argument that
there can be no obstruction to the construction
of an exact solution.  Consider the problem first from
the conformal field theory point of view.
The
string propagation in a given background metric is described, for
large radius, by a $\sigma$-model,
\beq
\int d^2 \sigma g_{i \bar j}(x^k,x^{\bar k})
\partial x^i \bar \partial x^{\bar j} + {\rm fermionic~~ terms,~
etc.}
\label{sigmamodel}
\eeq
In this form, it is clear that the expansion parameter is $R^2$,
where $g_{i \bar j} = R^2 g^{o}_{i \bar j}$, where $g^{o}$ is
a reference metric of order the string scale.  To lowest order in
$R^2$, the condition for conformal invariance is vanishing of the
$\beta$-function,
\beq
\beta_{i \bar j} = {\cal R}_{i \bar j} =0
\eeq
where ${\cal R}$ is the Ricci tensor.

The conformal field
theories which describe Calabi-Yau backgrounds have two left-moving and
two right-moving supersymmetries.  The question of obstructions,
here, is whether the conformal field theory constructed as a solution
of the lowest order equations generalizes to a solution to all
orders.  Tadpoles for massive fields correspond to corrections to
the $\beta$-functions of irrelevant operators, and are not an
issue; the question is whether there are corrections to the
$\beta$-functions for the marginal operators.  To see that
there cannot be, note that these
are suitable backgrounds
not just for the heterotic string theory
but for the Type II theory;
the Type II theories have both left and right
moving supersymmetries.  In the Type II theory, in space-time,
these solutions describe backgrounds with $N=2$ supersymmetry.
But we have seen that for $N=2$ supersymmetry, the moduli cannot
be exact.  This implies that the $\sigma$-model is an exact
conformal theory,
and this remains true whether it is considered as a background for the
heterotic theory or Type II theory.
These statements are supported by detailed perturbative
computations, as well as by analyses of instanton effects in the
$\sigma$-model.

In situations with less world-sheet supersymmetry, the situation
is more complex.  To understand the nature of the problem, we can
give another argument, more directly in the heterotic string.
Among the various moduli of the theory are deformations of the
metric, $\delta g_{i \bar j}$ of eqn. [\ref{sigmamodel}] which preserve
the conformal-invariance condition.  These are in one to one
correspondence with ($1,1$) forms, $b_{i,\bar j}$,
\beq
b_{i \bar j} = \delta g_{i \bar j} = -b_{\bar j i}.
\eeq
The simplest example is $\delta g \propto g$, corresponding to an
overall dilation of the metric; this is clearly a solution, at
lowest order, since the ${\cal R}=0$ condition does not determine
the radius of the compact space.  In general, the Ricci-flat
metrics of interest are Kahler,
\beq
g_{i \bar j} = {\partial \over \partial x^i} {\partial \over
\partial x^{\bar j}}K.
\eeq
The vertex operator for $b$ is
\beq
V_b= \int d^2 \sigma {\partial \over \partial x^i} {\partial \over
\partial x^{\bar j}}K(\partial x^i \bar \partial x^{\bar j}
- \partial x^{\bar j} \partial x^i)e^{i k \cdot x} + {\rm terms~~
which ~~vanish~~ at}~~ k^{\mu}=0.
\eeq
Here $k$ is the ordinary momentum four vector, and $x^{\mu}$
refers to the free fields which describe the non-compact
dimensions.
At zero momentum, the integrand is a total derivative,
\beq
V_b = \int d^2 \sigma \partial({\partial K \over \partial x^{\bar
j}}\bar \partial x^{j}) - \bar \partial({\partial K \over \partial x^i} \partial x^{\bar i})
.
\eeq

In $\sigma$-model perturbation theory, then,
this mode of the metric decouples at zero momentum.
This means that the effective four-dimensional theory has a
symmetry under which $b$ shifts by a constant.  $b$ is part of a
chiral multiplet.  In the case of the radial dilaton, the scalar
component of this multiplet is $R^2 + ib$.  But $W$ is an analytic
function of this field; because it is independent of $b$ it is
also independent of $R^2$.  In other words, there are no
corrections to the superpotential in $\sigma$-model perturbation
theory!

These statements hold for both the ``standard embedding," with
spin connection equal to the gauge field, and more generally.  One
can also ask what happens beyond perturbation theory.  We have
already given an argument that for the standard embedding there
are no non-perturbative corrections either.  This can be verified
at the level of world-sheet instantons.  For more general
compactifications, it appears that generically there should be
corrections.  The actual situation is more complicated,
however\cite{silverstein}.

Arguments of this type can also be used to establish the existence
of massless states other than moduli, not protected by any
space-time symmetry.  This could well be relevant to the MSSM, where
one wants to understand the absence (at the level of the
superpotential) of a mass for the Higgs fields.

\subsection{Beyond the Classical Approximation}

Using space-time arguments, there is a good deal one can say about
these theories non-perturbatively.  In the Type II case, $N=2$
supersymmetry again ensures that there is no potential for the
moduli.  In the heterotic case one has only $N=1$ supersymmetry.
In this theory, the dilaton ($1/g^2$) is in a multiplet with the
axion, $B_{\mu \nu} \leftrightarrow a$,
\beq
\int d^2 \theta[{1 \over g^2}+ia] W_{\alpha}^2
\eeq
$$~~~~~~~=\int d^2 \theta S W_{\alpha}^2.$$
Again, there is a shift symmetry, $a \rightarrow a + \delta$.
This can be understood by looking at vertex operators, or by
noting that $\partial a \propto ~~^*H$, where $*H$
denotes the dual of $H$ ($*H_{\mu}
= {1 \over 4!} \epsilon_{\mu \nu \rho \sigma}
H_{\nu \rho \sigma}$), and that the $B$ gauge
symmetry forbids non-derivative terms for $H$.

So, in string perturbation theory, since $W$ is holomorphic, there
are no dilaton dependent corrections to the superpotential.
Moduli remain moduli, massless particles remain massless.  What
happens beyond perturbation theory?  The axion shift symmetry is
anomalous.  Instantons of the low energy field theory, for
example, can generate effects which behave as $e^{-S}=e^{-
8\pi^2/g^2 + ia}.$  Non-perturbative, stringy effects surely
generate similar terms.  Given our lack of understanding of the
theory at a fundamental level, one might think that there is
little one can say.  However, it turns out that using symmetries
and holomorphy, it is often possible to make striking statements.

First, we should note that we do not expect that in a theory of
gravity there should be continuous global symmetries.  In weakly coupled
string theory, this is a theorem\cite{globalsymmetries}.
In all of the recent work on strongly coupled theories, no
examples of continuous global symmetries have been found.  We will
thus adopt as a working hypothesis that the only allowed
continuous
symmetries are gauge symmetries.

Discrete symmetries, on the other hand, abound in string theory.
They can usually (and probably always) be thought of as gauge
symmetries.  It is easy to construct examples.  In toroidal
compactification on a square lattice there is a symmetry which
interchanges the two lattice generators.  This $Z_2$ symmetry
traces back to general coordinate transformations in the compact
space; it is an $R$ symmetry, since it rotates fermions
differently than bosons.  Much more intricate versions of
such symmetries are discussed in \cite{gsw} in the
case of Calabi-Yau compactification.  Again, generically
these transformations, which originate as coordinate transformations
in the internal dimensions, act differently on fermions than bosons,
and so are $R$-symmetries.  The $Z_2$ which exchanges the
two $E_8$'s of the heterotic string theory is another discrete
symmetry (it can also be shown to be a gauge symmetry).

Such discrete symmetries place powerful constraints on the form of
the low energy theory.  As an example, consider the (Wilsonian)
effective action at some high energy scale for a compactification
of the heterotic string with some set of discrete $R$ symmetries.
Typically, the dilaton, $S$, is neutral. On the other hand, the
superpotential must transform by a phase under the various $R$
symmetries.  So \beq \int d^2 \theta f(e^{-S}). \eeq Provided that
there are
not couplings of the type ${\cal M} e^{-S}$, for some other
modulus, ${\cal M}$, no perturbative or non-perturbative effects
can lift the flat directions.  In many cases,
such couplings are forbidden by symmetries.  Examples of this phenomena occur
already in textbook models, such as the quintic in $CP^4$ and the
$Z_3$ orbifold\cite{miracles}. Recall that this is a statement
about some Wilsonian action with a large cutoff.  What this says
is that SUSY breaking, if it occurs at all, must be a phenomenon
of the low energy field theory!

Let us focus, then, on this low energy field theory.  For example,
in the case of the standard embedding, the low energy gauge group
is $E_6 \times E_8$ (or a subgroup of $E_6$ obtained from Wilson
lines).  For the $E_8$, there are no matter fields.  As we have
seen, in such a theory, there is gluino condensation.   The
dilaton couples to the gauge fields through $\int d^2 \theta S
W_{\alpha}^2$, which leads to a superpotential for $S$
proportional to $\langle \lambda \lambda \rangle$.  One can, in
fact, determine the form of $W(S)$ completely from symmetries. The low
energy theory has an approximate symmetry under which
\beq
\lambda \rightarrow e^{i \alpha} \lambda ~~~~~~S \rightarrow
S - i {b_o \over 3} \alpha.
\eeq
To be consistent with this, $W$ must take the form
\beq
W = c e^{-{3S \over b_o}}.
\label{gluinocondensation}
\eeq

We have argued that the global symmetry cannot be exact, and we
can at this point ask about the form of possible corrections.
Suppose, for example, that in the high energy theory there is a
term
\beq
\zeta \int d^2 \theta W_{\alpha}^2 W_{\beta}^2.
\eeq
Such a term could arise at one loop.  Treat $\zeta$ as a spurion.
It then has $R=-2$, so one might expect corrections to the
superpotential of the form
\beq
\zeta e^{-{6S \over b_o}}.
\eeq
This correction is systematically smaller than that of eqn.
[\ref{gluinocondensation}] at weak coupling.

The main features of eqn. [\ref{gluinocondensation}] and its possible
corrections should not come as a big surprise.  In general, we
expect $V(S) \rightarrow 0$ as $S \rightarrow \infty$.  Similarly,
as $R \rightarrow 0$ (say with fixed value of the ten-dimensional
dilaton), the theory has more supersymmetry, and the potential
vanishes.  What is interesting is how much control we have over
the theory in these cases.

The fact that the potential tends to zero at weak coupling and
large radius means that stable minima of string theory exist, if
at all, at strong string coupling and compactification radii of
order one (except in those cases where the moduli are not lifted
at all).  This is the big puzzle of string phenomenology.  Why are
the gauge couplings we observe weak?  Why is $M_{gut} /M_p$ small?
And finally, is anything calculable in any kind of controlled
expansion?

Note that duality, by itself, is not much help in addressing these
questions.   {\it Any} weak coupling or large radius description
would lead to the same difficulty.  However, holomorphy, as we
will argue shortly, may offer, again, some hope for understanding.

\subsection{Alternatives to the Weak Coupling Viewpoint}

There are at least two troubling features of the weak coupling
picture of string theory.  The first is theoretical, the second
more phenomenological.  These are
\begin{itemize}
\item  The strong coupling problem which we have described above,
i.e. the fact that one can't stabilize the moduli in a systematic,
weak-coupling approximation.
\item
Even assuming the moduli are somehow stabilized at weak coupling,
there is a difficulty.  In the heterotic string, the four
dimensional coupling is related to the dimensionless string
coupling through
\beq
g_4^2 = g_s^2 ({l_s^2 \over V}).
\eeq
The four dimensional gauge couplings are numbers of order $1$
(typical unified couplings are about $1/\sqrt{2}$).  If the theory
is weakly coupled, $g_s \le 1$.  So one requires
$V/l_s^6 \sim 1$.  But this means that
$M_p \approx M_s \approx M_{GUT}$, since $M_{GUT} \approx R^{-1}$,
where $R$ is the compactification scale.  This is hard to
reconcile with the fact of perturbative unification.
Alternatively, if one takes $2 \times 10^{16}$ GeV as the
unification scale, one predicts that the string coupling is
gigantic!
\end{itemize}

One can object that perhaps one should consider weakly coupled
string theories other than
the heterotic string.  But difficulties arise in these cases as
well.  In particular, in the Type I theory,
\beq
G_N \propto {\ell_s^8 \over V} g^2~~~g_4^2 \sim 1 \sim {\ell_s^2
\over V} g
\eeq
so $g^2 \sim 10^{-12}$!  It is hard to imagine how the coupling
could be stabilized at such a small value, but this is not a
question which has been extensively explored.

\begin{figure}[htbp]
\centering
\centerline{\psfig{file=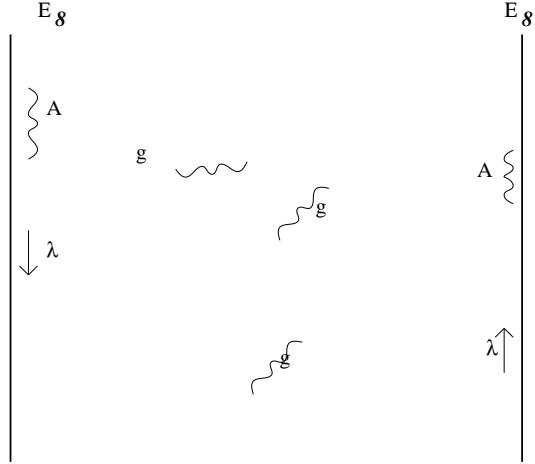,angle=-90,width=7cm}}
\caption{Structure of the strongly coupled heterotic theory.}
\label{horavawittenfig}
\end{figure}

A far more plausible picture emerges, if, following Horava and
Witten\cite{hv} and Witten\cite{wittency} we consider the strongly
coupled limit of the heterotic string.  In this limit the theory
looks eleven dimensional, with two walls.  Supergravity (the
graviton, three index antisymmetric tensor, and gravitino) live in
the bulk, and gauge fields live on the walls, as in fig. \ref{horavawittenfig}.
The
effective lagrangian looks like
\beq
{\cal L}= -{1 \over 2 \kappa^2}
\int_{M^{11}} d^{11} x \sqrt{g}( {\cal R} - \sum_{i=1}^2{1 \over 8
\pi (4 \pi \kappa^2)^{2/3}}) \int_{M_i^{10}} d^{10}x \sqrt{g}( {\rm
tr ~}
F_i^2 + \dots)
\label{hvaction}
\eeq
This action is useful only in the limit that all length scales
are large compared to the eleven dimensional Planck length,
$\ell_{11}$.

Now suppose that six dimensions are compactified, say on a
Calabi-Yau space.  The small parameters are  $1/R_{11}$,
where $R_{11}$ is the inverse separation of the
two walls (the size of the eleventh dimension), and
$1/V$, the volume of the Calabi-Yau space.
From eqn. [\ref{hvaction}] it follows that
\beq
M_{11} = R^{-1} (2(4 \pi)^{-2/3}\alpha_{GUT})^{-1/6}
~~~~~~R_{11} = {\alpha_{GUT}^3 V \over 512 \pi^4 G_N^2}.
\eeq
Plugging in numbers, one finds that $R \sim 2 M_{11}^{-1}$,
$M_{11} R_{11} \sim 72.$
So this is clearly a better picture than
the weak string coupling picture!  Both $1/R_{11}$ and $1/V$ are
reasonably small.

But now there is a puzzle, connected with how the moduli are
fixed, similar to those we discussed from a weak coupling string
viewpoint.  We can, again, attempt to construct a low energy
effective lagrangian for this theory.  Indeed, we can identify
fields similar to those we identified for weak coupling
compactifications.  As one can see from our formulas for the
couplings, $1/V \sim S$, the usual weak coupling dilaton.  $C_{11
\mu \nu} \sim a$, what is usually called the model-independent
axion.
$T \sim R_{11}R^2$ ($R$ here is the Calabi-Yau radius),
while $C_{11 ~I J} \sim b$.  The lagrangian is
constructed from $dC$, so it is invariant under shifts by harmonic
forms.  This is the same shift symmetry as in the weak coupling
theory.  Finally, just as at weak coupling, if the Calabi-Yau has
discrete $R$ symmetries (as for the quintic in $CP^4$ discussed in
\cite{gsw}), supersymmetry breaking must be a phenomenon of the
low energy theory.

The question of supersymmetry breaking and of moduli stabilization
is, in such cases, a question of whether the low energy theory
generates a superpotential.  One picture for how such breaking
might arise, which closely mirrors the weak coupling picture, is
to suppose that the interactions of the standard model reside on
one wall, while the other wall contains some additional, ``hidden
sector" gauge interactions which give rise to gluino condensation.
Gluino condensation, again, will generate a superpotential for a
linear combination of $S$ and $T$.
The superpotential one obtains agrees exactly, for a given
Calabi-Yau space, with the weak coupling result.  This is not
surprising.  The parameter for the
superpotential calculation is $e^{-S}$, $e^{-T}$.
It is possible for $S$ and $T$ to be large, while the
string coupling ($S^{-1}T^3$) is large.  So while the weak
coupling string theory and eleven dimensional supergravity limits
of the theory do not have overlapping regions of validity, the
superpotential calculations do.  It is thus important that they
agree, and supports not only the duality between these regimes but
the very existence of the theory with reduced supersymmetry.

To actually determine the potential requires knowledge of the
Kahler potential.  In the weak coupling limit, one could easily
read this off at tree level.  At strong coupling, the situation is
similar.  At large $R_{11}$, the theory goes over to a five
dimensional, $N=1$ theory, i.e. a theory with eight
supersymmetries.  These symmetries are highly restrictive.  In
particular, the Kahler potential is necessarily the same at strong
as at weak coupling.  As a result, just as at weak coupling, one
has a potential,
$$V \sim {1 \over R_{11}^6} e^{-aR_{11} -S}.$$
Here, the term $a R_{11}$ in the exponent arises because the
gauge coupling in the hidden sector depends not only on $S$ but on
$T$.  This is true both at weak coupling\cite{nillesetal,bdhv} and
strong coupling\cite{wittency}.  Again, this coupling, which is
holomorphic, agrees in the two limits\cite{bdhv}.

We see, as we might have expected, that the moduli can be
stabilized, if at all, for $S$, $R_{11}\sim 1$.  So, while the
Horava-Witten limit certainly yields a better qualitative
description of the theory than does the weakly coupled string, but
the stabilization issue remains.

\section{Moduli and Cosmology}

\begin{figure}[htbp]
\centering
\centerline{\psfig{file=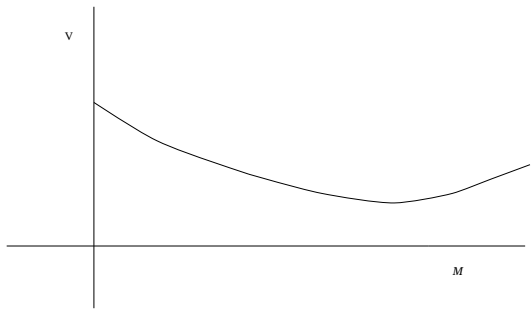,angle=-90,width=7cm}}
\caption{A plausible potential for the moduli.}
\label{moduluspotential}
\end{figure}

A possible clue to understanding the fate of the moduli of string
theory comes from cosmology.  Suppose that there is a stable
vacuum of string theory, with broken supersymmetry and in which
the moduli gain mass of order $m_{susy}$.  One might imagine that
the modulus potential looks as in fig. \ref{moduluspotential}.  Such a picture has
two possible implications for cosmology:
\begin{itemize}
\item
The potentials for the moduli are likely to be rather flat.  Thus
moduli are candidate inflatons\cite{bkn,banksinflation}.
\item
The moduli have the potential to carry far too much energy,
overclosing the universe\cite{bkn}. This is called the ``moduli
problem" of string cosmology.  Related problems include the
question of whether the system can even find the correct
vacuum\cite{bs}.
\end{itemize}

Through much of these lectures, we have stressed the possibility
that the vacuum which describes our universe lies in an
approximate moduli space.  If one assumes these moduli have masses
of order $M_W$ (or smaller), and that they range over values of
order $M_p$, one encounters a serious cosmological difficulty.
The fluctuations of the microwave background
suggest that the Hubble parameter, $H \sim {1 \over t}$, was once
of order $10^{16} ~{\rm GeV}
\gg M_W.$  At such times, effects other than the zero temperature,
low curvature, effective action were probably important.  For
example, if there was an earlier period of inflation, associated
with an
inflaton, one would expect large corrections to the potential for
the scalar fields.  Calling ${\cal I}$ the inflaton, and assuming
that the underlying laws are supersymmetric, one might expect
${\partial W \over \partial {\cal I}} \sim M_p H
\sim (10^{16} ~{\rm GeV})^2$.  Then couplings such as
\beq
{1 \over M_p^2} \int d^4 \theta {\cal I}^{\dagger} {\cal I}
{\cal M}^{\dagger}{\cal M}
\eeq
give big corrections to the moduli potential.  Generically, there
is no reason that ${\cal M}$ should sit at the minimum of its
potential at these early times.  Indeed, until $H\sim M_W$, the
zero temperature, zero curvature potential is presumably
irrelevant.  At $H\sim M_W$, then, it is natural to suppose that
${\cal M}-{\cal M_o} \sim M_p$.  The equation of motion for
${\cal M}$ is just $D_t^2 {\cal M} + V^{\prime}$, which
in the Robertson-Walker
background becomes:
\beq
{\cal M} + 3H \dot {\cal M} + V^{\prime} ({\cal M})=0.
\label{mevolution}
\eeq
Assuming that the universe is matter dominated at this time (e.g.
due to the oscillations of the inflaton field),
\beq
H={2\over 3t}
\eeq
corresponding to a scale factor growing as $R \sim t^{2/3}$.
Eqn. [\ref{mevolution}] then has the solution
\beq
{\cal M} \approx {\cal M}_o \sin(mt)({t_o \over t}).
\eeq
where we have written the quadratic term in potential near the
minimum as
$m^2({\cal M}-{\cal M}_o)^2$.  The energy is approximately
\beq
\rho = {1 \over 2}[\dot {\cal M}^2 + m^2 {\cal M}^2]
\propto ({R_o \over R})^3.
\eeq
One can think of
this as a coherent state of massive particles, oscillating in
phase, and diluted by the expansion of the universe.

\noindent
{\bf Exercise:}  Show that the same is true in the radiation
dominated case, i.e. when $R \propto t^{1/2}$.

The difficulty is that these particles typically come to dominate
the energy density of the universe long before nucleosynthesis.
One expects that they have Planck scale couplings, so their
lifetimes are not likely to be shorter than
\beq
\Gamma \le {1 \over 4 \pi^2}{m^3 \over M_p^2} \sim 10^{-30}
\eeq
for TeV mass moduli.  This corresponds to a lifetime of order
$10^7$ sec, i.e. much later than typical times associated with
nucleosynthesis.  The density at the time of decay can be
estimated as follows.  By assumption, ${\cal M}$ starts at a
Planck distance (in field space) from its minimum.  It starts to
oscillate when the Hubble constant is comparable to its mass, i.e.
when
$t_o \sim {1 \over m}$.  Thus initially the energy density
is $\rho \sim m^2 M_p^2$, so when the moduli decay,
\beq
\rho = m^2 M_p^2 ({\Gamma \over m})^2.
\eeq
Taking $m=10^3 ~{\rm GeV}$, this is about $10^{-20}$ GeV.
Thus when the decay products thermalize, their temperature will be
of order $1 {\rm KeV}$.  This is well below the temperature of
nucleosynthesis.  The successful predictions of nucleosynthesis
are thus spoiled (these are based on the assumption that the
universe is radiation dominated at nucleosynthesis; moreover,
the decay of the moduli destroys most of the light
nuclei).

This problem is called the moduli problem of string
cosmology.  Various solutions have been proposed
\begin{itemize}
\item
The moduli masses are much larger than one TeV.  If the mass is
$100$ TeV or so, the reheat temperature is higher than a few MeV,
and nucleosynthesis can occur again.  The possible difficulty here
is that one must also produce the baryons in the decays of these
particles.  This might occur through baryon number violating
couplings, or through the Affleck-Dine mechanism\cite{ad}
\item
Late inflation\cite{rt}.  The idea here is that a late stage of
inflation drives the moduli to their minima.  The difficulty with
this proposal is that the minima, as we argued above, will not
necessarily coincide with the zero curvature minima, so this
possibility seems to be fine tuned.
\item
Enhanced symmetries.  We have seen that sometimes in string
theory, all moduli are charged under unbroken symmetries at some
points in the moduli space.  These points of ``Maximally enhanced
symmetry" are automatically stationary points of the full quantum
effective action.  They are also naturally minima at early times,
so it is plausible that the high curvature (temperature) and zero
curvature (temperature) minima coincide.
\item
There are no approximate moduli.   This possibility, in some ways,
is not so different than the previous one.   It obviously avoids
the moduli problem.  But it has the inherent problem that there is
not likely to be a small parameter in such a picture.  Moreover,
as we have discussed above, it is unclear how one would connect
such a possibility to string theory.
\end{itemize}

There are other troubling issues in string cosmology connected
with moduli, which we do not have time to review here\cite{bs,
hm,banksberkooz}.  But for
the moment, we would argue

\section{Stabilization of Moduli}

We turn, then, to the question of stabilization of the moduli.  No
complete model exists, but there are some ideas.  Given the poor
state of our present understanding, we should try, as we review
these ideas, to keep in mind certain questions:
\begin{itemize}
\item  Are there generic predictions in any particular scheme?
For example, is there low energy supersymmetry with some pattern
of soft breakings?
\item  Why are the gauge couplings weak?  Why are the radii large?
\item  What quantities, if any, are calculable, even in principle?
\end{itemize}

We will discuss four proposals which have been put forward from
this viewpoint:  Kahler stabilization, the racetrack scheme,
maximal symmetry, and topological stabilization in large
dimensions.

\subsection{Kahler stabilization}

In weak coupling, the string perturbation expansion is believed to
be less convergent than that of ordinary field theory.  This
motivates the hypothesis that the Kahler potential of the moduli,
$K({M})$, is much different from its weak coupling form when
$e^{-S}$ ($S$ is the modulus which determines the four dimensional
gauge couplings, which we will call the dilaton).  Indeed, this is
compatible with our discussion of the Horava-Witten scheme, where
we saw that there is a regime of large $S$ and $R_{11}$ where the
weak coupling picture is certainly not valid.  (In that limit, the
question is to understand why the corrections to the $T$ Kahler
potential are so large in a regime where $T$ itself would seem to
be large).  Focusing on the dilaton, suppose
$W = e^{-cS}$
as in models of gaugino condensation.  One can easily invent
Kahler potentials
such that the potential of eqn. [\ref{supergravity}] has a minimum at
some point $S_o$ such that $e^{-S_o}$ is
hierarchically small and with vanishing cosmological
constant\cite{kahlerpotentials}.  These models, of course, are
finely tuned, but at least the various terms in the potential are
comparable in order of magnitude.

What predictions does this hypothesis make?
Beyond the starting assumption that the theory is approximately
supersymmetric (i.e. low energy supersymmetry), there are two
generic outputs of this scheme:
\begin{itemize}
\item  Coupling unification:  if the couplings are unified for
large $S$, as in weakly coupled strings,
\beq
\int d^2 \theta (S + e^{-S} + \dots) W_{\alpha}^2
\eeq
is hardly corrected from its large $S$ form.  Note, however, that
in this scheme one has no control over threshold corrections.   In
practice, this means that at most one can trust the leading log
contributions to coupling unification.
\item  For similar reasons, the terms in the superpotential are
the same as in the large $S$ limit, for similar reasons as above.
\end{itemize}

However, in this scheme, the Kahler potential, by assumption, is
modified from its large $S$ form.  So quantities like soft
breaking masses are inherently uncomputable.

\subsection{Racetrack Models}

Suppose one has two gauge groups without matter fields, each of
whose couplings is determined by a modulus $S$.  Then the
superpotential has the form
\beq
W = b_1 \alpha e^{-S/b_1} -b_2 \beta e^{-S/b_2}.
\eeq
The equation
${\partial W \over \partial S}=0$
yields
\beq
S= {b_1 b_2 \over b_2-b_1} \ln(\beta/\alpha).
\eeq
If $b_1$ and $b_2$ are large, then $S$ is large.  But unless
$b_1 \approx b_2$, $e^{-S/b_1} \sim e^{-S/b_2} \sim 1$,
and the low energy analysis is inappropriate.  So one requires
$b_1 = b_2 + \delta$.  This represents a discrete fine tuning.
For example, if the gauge groups are $SU(10)$ and $SU(11)$,
$S \approx 100$, which is not too far from a grand unified
coupling.

In supergravity, this minimum yields a state with a non-zero
cosmological constant.  This problem can be solved, however,
if the model has an $R$ symmetry, unbroken at the
minimum\cite{yanagida}.  The one existing model requires many singlets, with
constraints on their couplings which presumably require rather
elaborate discrete symmetries.  One might hope to find a more
economical model.

\begin{figure}[htbp]
\centering
\centerline{\psfig{file=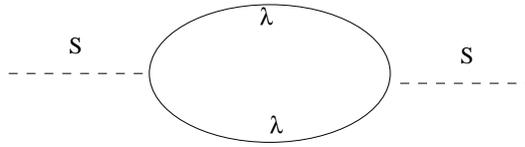,angle=-90,width=7cm}}
\caption{Large correction to the Kahler potential in racetrack
models.}
\label{kahlerracetrack}
\end{figure}

While this scheme might generate a small gauge coupling with unbroken
supersymmetry and one less modulus,
there may be no sense in which there is a weak coupling expansion.
Consider, for example, the corrections to the $S$ Kahler
potential, indicated in fig. \ref{kahlerracetrack}, from the
point of view of the heterotic string theory.  Because $N^2$ gauge fields run
in the loop, diagrams like this and its generalizations behave
as $(g^2 N^2)^m$, i.e. the would-be expansion parameter is of
order one.  Still, as for Kahler stabilization, holomorphy can
ameliorate this problem.  $e^{-S/b_1} \ll 1$.  So holomorphic
quantities computed at large $S$ maintain there form even at the
would-be minimum.  In the Type I theory, the situation is somewhat
different (I thank I Antoniadis for comments
which lead to consideration of this point.)
Since the string coupling, $g$, is equal to
$g_{YM}^2$, loop corrections might be small (modulus factors
can also enter in the heterotic string argument above).
It is not completely clear what one should make of this latter
observation, since in any case there is no sense in which one can
take the gauge group arbitrarily large.  We will assume, for the
rest of this section, that in the racetrack picture with a
discrete fine tuning, one can compute holomorphic but probably
not non-holomorphic quantities.

The result is not so unappealing.  One discrete fine tuning is the
main thing which is required, to fix the ``dilaton."  Gauge
couplings may be calculable.  The mass of the dynamics which fixes
the modulus is larger than the scale of supersymmetry breaking.
Supersymmetry breaking itself should be through low energy
dynamics.  Physical quantities related to holomorphic quantities
should be calculable.

\subsection{Maximally Enhanced Symmetry}

We have seen that there exist points in moduli spaces at which all
moduli are charged under symmetries.   The particular examples we
discussed had at least $16$ supersymmetries.  One can find an
example with $8$ supersymmetries (N=2 in four dimensions) by
considering the Type II theory on a Calabi-Yau space at a ``Gepner
Point."  At the Gepner point, all of the geometrical moduli are
charged under symmetries.  The theory inherits the $SL(2,Z)$
duality of the higher dimensional theory, so by going to the
self-dual point, one has a situation where all moduli transform
under symmetries.

What about $N=1$?  Such points presumably exist, and they are
interesting from at least two points of view:
\begin{itemize}
\item
They are naturally stationary points of the effective action.
\item
If the ground state of the system is such a point, this provides a
natural solution of the cosmological moduli problem.
\end{itemize}

The main objections to such a possibility are that one expects
$\alpha \sim 1$, as in the self-dual point of electric-magnetic
duality, and that one does not expect hierarchies, nor that any
sort of systematic approximation schemes should be available.
Still, we can adopt the hypothesis that there exist such states,
with gauge couplings unified and
$\alpha \ll 1$, and with $e^{-2 \pi/\alpha}$ small, allowing for
hierarchies.  This is at least consistent with the observed facts,
has the two virtues listed above, and has some definite
consequences:
\begin{itemize}
\item
SUSY breaking is a low energy phenomenon.  This is because the
symmetries forbid terms in the superpotential linear in any of the
moduli.
\item $\int d^2 \theta {\cal M} W_{\alpha}^2$
is also forbidden, so gluino condensation, in the conventional
sense, is also irrelevant for supersymmetry breaking
\item  Related to the previous point,
gaugino
masses cannot be generated by moduli $F$-terms.  Thus it is
probably necessary that supersymmetry be broken as in gauge
mediated models at quite low energy.
\end{itemize}

An interesting observation concerns the identity of the moduli in
this picture.  They could well be fields of the MSSM.  For
example, at the level of renormalizable interactions, there is a
flat direction in the MSSM with
\beq
Q^{[1]} = \left ( \matrix {v & & \cr & v & \cr & & 0}
\right )~~~~~Q^{[2]}=  \left ( \matrix {0 & & \cr & 0 & \cr & & v}
\right )
~~~~~~~L= \left ( \matrix{0 \cr v} \right ).
\eeq
Here the numbers in braces are $SU(2)$ indices.
This flat direction has a gauge invariant description in terms of
the chiral field $QQQL$.  This direction can be exactly flat if
the fields transform suitably under $R$ symmetries.  So the moduli
might well be superpartners of known particles, exactly flat due
to discrete $R$ symmetries.

Two related possibilities should be mentioned.  First, it could be
that there are no moduli, even in some approximate sense.
Operationally, this is not too much different than the maximal
symmetry hypothesis.  Again, there is the puzzle:  why are the
couplings small?  Why are there hierarchies?   One can also
consider the racetrack model in combination with enhanced
symmetries.  The modulus which controls the gauge coupling might
be fixed by supersymmetry-conserving dynamics, and very
massive, with the rest of the moduli
sitting at enhanced symmetry points.

\section{Large Dimensions and TeV Scale Strings}

It has usually been assumed that the compactification scale of
string theory should be similar to the string scale (or the eleven
dimensional Planck scale, etc.).  The underlying
prejudice is simply that dimensionless ratios shouldn't be
terribly large.  In the heterotic string, one can make this idea a
bit more precise, by noting, as we did earlier, that weak string
coupling, and the observed values of the gauge couplings, imply
that the scales cannot be too different.  Similarly, in the
Horava-Witten description, none of the scales are wildly
different.

Over the past year, however, there has been much interest in a
different possibility.  It has been argued that perhaps the
solution to the hierarchy problem lies not in supersymmetry, but
rather in the possibility that the fundamental scale of
interactions is roughly $1$ TeV\cite{largeradii}.  Much as in the Horava-Witten
picture, the Planck scale is a derived quantity, large only
because some internal dimensions are now (very) large.  The
standard model fields should live on a brane or wall, so that the
standard model couplings do not become extremely small as the
volume of the internal space tends to infinity.  The picture,
then,
is essentially as in fig. \ref{horavawittenfig}, with the understanding
that the fields on the branes are the standard model fields, and
perhaps those associated with additional interactions.

How large is the internal space?  This depends on the number of
compact dimensions.  Suppose, for definiteness, that the
underlying theory is ten dimensional.  Suppose that there are six
compact dimensions.  The four dimensional Newton's constant is
given by
\beq
G_N = {(\ell_{10})^8 \over V_6} = {1 \over M_p}^2 ~~~~~l_{10}
\sim ~{\rm TeV}^{-1}.
\eeq
Then if there are six large dimensions of comparable size, $r$,
$r \approx ({\rm MeV})^{-1}$, while if there are two large
dimensions, with the others of size $l_{10}$, $r \approx {\rm
mm}$!

This latter possibility is quite amazing, and even more surprising
it is not so easy to rule out.  In particular, it means that if
one probes distances shorter than a millimeter, one will see a
modification of Newton's laws, appropriate to five spatial
dimensions rather than three, i.e.
\beq
F \sim {1 \over r^4}
\eeq
This is a dramatic prediction, just barely compatible with current
experimental limits, and accessible to improved experiments.
As we will see shortly, however, there are astrophysical constraints which
suggest that the scale, in the case of two large dimensions,
must be significantly larger, probably
placing the modification of Newton's law out of reach.

For any number of compact dimensions, however, there are also
dramatic effects in high energy collisions as one approaches the
TeV scale.  At these energies, one ``sees" the extra dimensions.
This is because, while each Kaluza-Klein state has coupling of
order $1/M_p$, there are {\it many} states.  The fact that
individual states couple with strength $1/M_p$ follows from the
form of the terms in the lagrangian
\beq
\ell_{10}^{-8} \int d^6 y d^4 x \sqrt{g} {\cal R} = M_p^2 \int d^4 x
\sqrt{g} {\cal R},
\eeq
so the coupling of each Kaluza-Klein mode is suppressed by a
factor of the volume.  However, there are lots of modes; once $E
\gg R^{-1}$, one has a phase space integral appropriate to the
higher dimensional field theory, i.e. amplitudes behave like
\beq
V_6 \int {d^d k \over (2\pi)^d} G_N \times {\rm kinematic factors}
\eeq
$$~~~~~~~\sim {1 \over {\rm TeV}^8} \int {d^d k \over (2 \pi)^d}
\times \dots.$$
In other words, the extra dimensions ``open up."

Other proposals have also emerged over the past year, including
a particularly interesting one by Randall and Sundrum\cite{rs},
in which the extra dimensions, in some sense, are not large,
but there are exponentially large differences in the metric on the
different branes.  These subjects are developing rapidly, and it
is not possible to review them here.  On the theoretical side,
there are any number of intriguing questions.  To solve the
hierarchy problem in this framework,
understand why the radii are large (or the equivalent statement in
the Randall-Sundrum picture).  In the large dimension case,
if one is to avoid introducing very small numbers by hand,
it seems necessary to have supersymmetry at least in the
bulk.  This supersymmetry seems to have little direct
consequence for low energy physics.  In the Randall-Sundrum
case, it seems possible to achieve hierarchies
without even bulk supersymmetry\cite{goldbergerwise}.
The exploration of these extreme regions of the (approximate)
moduli space is still in its infancy, and it is quite possible
that these issues will be better understood.

On a more phenomenological level, there are a number of issues
which any scheme in which the fundamental scale much below the
conventional unification scale.  These include
\begin{itemize}
\item  {\bf Proton decay}: In order to suppress proton decay to
acceptable levels, it is almost certainly necessary to have a
large discrete group.  For example, if
\beq
Q \rightarrow e^{2 \pi
i \over 12} Q~~~~~~~L \rightarrow e^{2 \pi i \over 12} L
\eeq
then the leading operator, of the form $L^3 Q^9$, has dimension
18, and the lifetime is of order
\beq
\Gamma = {m_p^{29} \over M^{29}} \times {\rm other~small~factors}
\eeq
where $m_p$ is the proton mass, and $M$ is the fundamental scale,
now supposed to be of order a few TeV.
\item {\bf Other problems of flavor:}  In order to resolve other
problems of flavor, such as rare decays, it is probably necessary
that there be additional approximate flavor symmetries.  A number
of authors have explored the possibility that there is indeed some
large (discrete) non-abelian flavor symmetry, perhaps
spontaneously broken on distant branes, both in order to
understand the absence of flavor violation, and to understand the
quark and lepton mass hierarchies\cite{largedflavor}.
\item{\bf Coupling Unification}:  There has been much work on the
question of coupling constant unification in this picture.  At
first sight, one might think that the usual field theoretic
analysis of unification is lost, but this is not quite true.  For
example, in the case of two large dimensions, the log of the large
mass scale is replaced by the log of the radius of
compactification.  Still, obtaining the supersymmetric unification
predictions is not generic, requiring, among other things, a
spectrum similar to that of the MSSM.  This is perhaps a bit troubling,
since it sounds as if one needs supersymmetry again as part of the
story.
\item {\bf Astrophysical Constraints:}  In the case of $d=2$,
there
are significant constraints from the supernova SN87a.  The problem
is that emission of Kaluza-Klein modes, in this case, can carry
off most of the energy.  This yields a constraint $M>50$
TeV\cite{supernova}.
Other tests, such as high precision electroweak measurements,
suggest that, more or less independent of $d$, the scale must be
larger than several TeV.
\end{itemize}

What these proposals have shown is that there are plausible
solutions to the hierarchy problem which do not involve
conventional low energy supersymmetry.  They have dramatic
consequences for experiment.  It is important to decide whether
any of these ideas (including low energy supersymmetry) is really
compatible with string theory -- or better, whether
any is a robust prediction of the theory.

\section{What is String Theory and How Would We Know It?}

It is not likely that, sometime soon, someone will simply
exhibit a solution of string theory with a spectrum
and interactions identical with what we see in nature.
What we should probably be striving for is a robust,
general prediction such as:
\begin{itemize}
\item
String theory predicts low energy supersymmetry
\item
String theory predicts large dimensions without low energy
supersymmetry
\item
String theory predicts a warped geometry, with large or infinite
dimensions, without low energy supersymmetry.
\end{itemize}

At the moment, we seem simultaneously close and far from these
goals.  We understand a great deal about supersymmetry in string
theory.  We also understand extreme regions of the moduli space
which give a brane picture and very large dimensions.
What is now crucial is to formulate some principle which
might select among these possibilities.  While it will be exciting
if experiments discover a new symmetry or new dimensions,
it would be wonderful if string theorists could commit
themselves beforehand to one or another (or some still unknown)
possibility.

Supersymmetry has been the focus of much
of these lectures.  Here, we have proposed some ideas of how
a complete picture might look.
perhaps we are in some approximate moduli space, and, while no
sort of weak coupling analysis is applicable, still, certain
quantities can be computed starting from a weak coupling
approximation.  We have argued that there are some experimental
hints that this is the case:  the existence of hierarchies, the
smallness of the gauge couplings, and their unification.  I am
optimistic that we can go farther, perhaps even before our
experimental colleagues discover -- or fail to discover --
supersymmetry at the Tevatron and LHC.
Much effort is being devoted at the present time to fleshing
out the large/warped dimension pictures in a similar fashion.

\noindent
{\bf Acknowledgements:}

\noindent
I thank Josh Gray for a careful reading of the manuscript.
This work supported in part by a grant from the U.S.
Department of Energy.

%%%%%%%%%%%%%%%%%%%%%%%%%%%%%%
%  Bibliography
%%%%%%%%%%%%%%%%%%%%%%%%%%%%%%

\end{document}